\documentclass[aps,pra,showpacs,superscriptaddress, twocolumn]{revtex4}
\usepackage{graphicx}
\usepackage{bm}
\usepackage{amsmath}
\usepackage{hyperref}
\usepackage[latin1]{inputenc}
\hypersetup{
     colorlinks=true,      
    linkcolor=blue,        
    citecolor=blue,        
    filecolor=magenta, 
    urlcolor=black          
}

\providecommand{\moy}[1]{\langle #1 \rangle}
\providecommand{\bra}[1]{\langle #1 \rvert}
\providecommand{\ket}[1]{\lvert #1 \rangle}
\providecommand{\braket}[2]{\langle #1 \rvert #2 \rangle}

\newcommand{\expec}[1]{\langle #1\rangle}

\usepackage{mathbbol}
\usepackage{cancel}
\begin{document}

 \title{Quantum information processing in phase space: A modular variables approach} 
 
\author{A. Ketterer}\email{andreas.ketterer@univ-paris-diderot.fr}
\affiliation{Laboratoire Mat\'eriaux et Ph\'enom\`enes Quantiques, Sorbonne Paris Cit\'e, Universit\'e Paris Diderot, CNRS UMR 7162, 75013 Paris, France}
\author{A. Keller}
\affiliation{Institut de Sciences Mol\'eculaires d'Orsay, B\^atiment 350, Universit\'e Paris-Sud, Universit\'e Paris-Saclay, CNRS UMR 8214, 91405 Orsay, France}
\author{S. P. Walborn}
\affiliation{Instituto de F\'{\i}sica, Universidade Federal do Rio de Janeiro, Caixa Postal 68528, 21941-972 Rio de Janeiro, RJ, Brazil}
\author{T. Coudreau}
\author{P. Milman}\email{perola.milman@univ-paris-diderot.fr}
\affiliation{Laboratoire Mat\'eriaux et Ph\'enom\`enes Quantiques, Sorbonne Paris Cit\'e, Universit\'e Paris Diderot, CNRS UMR 7162, 75013 Paris, France}
\date{\today} 

\begin{abstract}
Binary quantum information can be fault tolerantly  encoded in states defined in infinite dimensional Hilbert spaces. Such states define a computational basis, and permit a perfect equivalence between continuous and discrete universal operations. The drawback of this encoding is that the corresponding logical states are unphysical, meaning infinitely localized in phase space. 
We use the modular variables formalism to show that, in a number of protocols relevant for quantum information and for the realization of fundamental tests of quantum mechanics, it is possible to loosen the requirements on the logical subspace without jeopardizing their usefulness or their successful implementation. Such protocols involve measurements of appropriately chosen modular variables that permit the readout of the encoded discrete quantum information from the corresponding logical states. Finally, we demonstrate the experimental feasibility of our approach by applying it to the transverse degrees of freedom of single photons.

\end{abstract}
 
\maketitle

\section{Introduction} Following classical protocols, quantum information protocols were initially formulated in terms of {qubits} that are manipulated unitarily in order to realize computational and communication tasks that may over perform their classical analogs \cite{Nielsen}. A different, but widespread approach to process quantum information involves using continuous variables (CV) \cite{Loock}. Among the most important advances in the field of CV quantum information are the realization of quantum teleportation \cite{Braunstein, Furusawa1}, as well as quantum cryptography protocols, which rely on states defined in a continuous variables representation \cite{Grosshans1, Grosshans2}. Universality for manipulation of continuous variables quantum states was defined in Ref.~\cite{Loyd}, and subsequently measurement based quantum computation was generalized from the discrete to the continuous realm \cite{MBQCprl,MBQC}.  

Manipulating quantum information in continuous or discrete variables, on its usual circuit based formulation, involves the application of unitary gates. While such gates, in the discrete variable (DV) case can be expressed in terms of SU(2) transformations, a general unitary in CV is composed by polynomials of conjugate operators with a continuous unbounded spectrum, such as position and momentum of a particle, or the electromagnetic field quadratures. 
Even though, in the general case, a direct correspondence between universal operations in DV and CV has not been established, it was shown by Gottesman, Kitaev and Preskill (GKP) in Ref.~\cite{Gottesman}, that such a correspondence exists for a family of states that, while being defined in CV, can be used to encode a qubit. 
Moreover, the GKP encoding is also at the heart of the demonstration showing that fault-tolerant measurement-based quantum computation with CV cluster states is possible \cite{MenicucciFault}. 

A drawback of this encoding is that it relies on non-physical states. Furthermore, physical states that are close to them are of extremely challenging experimental realization with optical field quadratures. In addition, the GKP encoding has the specific purpose of quantum computation applications. In the more general context of quantum mechanics, quantum computing presents the particular aspect of requiring measurements realized in the computational basis only. However, a number of important quantum mechanics or quantum information related tasks, such as Bell inequalities violation \cite{CHSH}, quantum state tomography \cite{CVTomo}, and fundamental tests of quantum mechanics \cite{LG,CabelloContext1}, rather rely on the recovery of  binary information through the measurement of different mutually unbiased bases. For these applications, one should build a formalism offering an analogy of Pauli matrices in phase space not only from the operational point of view, as proposed in \cite{Gottesman}, but also from a measurable perspective.  Moreover, in order to build a complete toolbox to manipulate and measure discrete quantum information encoded in CV, one should also define how to perform rotations between different measurement bases.

In the present article we create a framework to manipulate and measure binary quantum information encoded in continuous variables using the formalism of modular variables (MV) \cite{Aharonov}, which allows us to naturally identify discrete structures in continuous variable states. We further introduce adapted operations and observables which enable us to manipulate and readout the encoded discrete quantum information in terms of the corresponding CV logical states. Our  formulation shows that, if one is interested in recovering quantum information by measuring binary observables defined in CV, one can loosen the requirements imposed on the GKP states. Our results have immediate experimental impact which we demonstrate by applying them to the transverse degrees of freedom of single photons.

The structure of this paper is as follows. In the next section we give an introduction to the modular variables formalism, including the definition of the modular position and momentum operators and the resulting representation in terms of their common eigenstates. In Sec.~\ref{sec:QIPframework} we present our main results and show how to process discrete quantum information encoded in logical states expressed in the modular variables representation. Further on, measurements of judiciously chosen modular variables are revealed to enable the readout the encoded discrete quantum information from the corresponding logical states. Section~\ref{sec:ExpProposal} is devoted to the discussion of an experimental implementation of our ideas using the transverse degrees of freedom of single photons. Finally, we conclude in Sec.~\ref{sec:Conclusion}.

\section{Modular variables formalism} \label{sec:ModVarForm}
\subsection{Definition of modular variables and the modular representation}
\begin{figure}[t]
\includegraphics[scale=0.5]{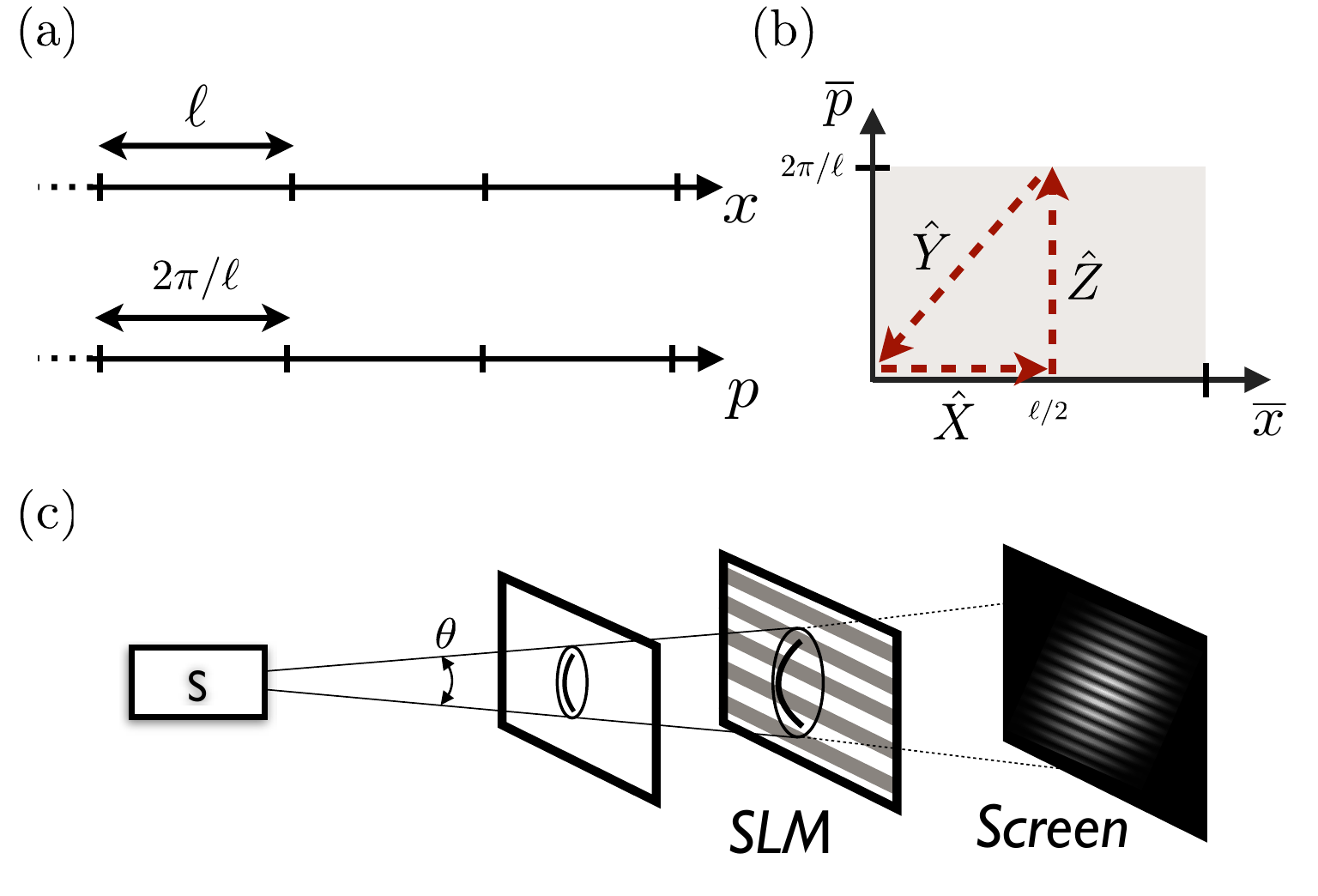}
\caption{ Schematic representation of (a) the unbounded position and momentum eigenspectrum together with  their discretization into boxes of length $\ell$ and $2\pi/\ell$, respectively, and of (b) the bounded spectra of the modular position and momentum operator. The red arrows indicate the displacements implementing the logical Pauli operators $\hat X$, $\hat Y$, and $\hat Z$. (c) Representation of the transverse distribution of a single photon created in a source S with a Gaussian wave function which is transformed into a periodic diffraction pattern by passing through a grating with slit distance $L$. In experiments, such a diffraction grating can be implemented easily using a spatial light modulator (SLM).}
\label{fig_1}
\end{figure}

In the modular variables (MV) formalism, pairs of canonically conjugate observables are expressed in terms of modular and integer parts, respectively. In the case of the position and momentum operators this leads to \cite{Aharonov}:
\begin{eqnarray}
 \hat{x} = \hat N {\ell}+ \hat{\bar x}, \hspace{1.5cm} \hat p = \hat M \frac{2\pi}{\ell}+ \hat{\bar p},
\label{eq:DefModVar}
\end{eqnarray}
where $\hbar=1$, $\hat N$ ($\hat M$) has integer eigenvalues, and $\hat{\bar x}=(\hat x+\ell/4) \text{mod}[\ell]-\ell/4$ ($\hat{\bar p}=(\hat p+\pi/\ell) \text{mod} [2\pi/\ell]-\pi/\ell$) is the {modular position} ({momentum}) operator with eigenvalues in the interval $[-\ell/4,3\ell/4 [$ ($[-\pi/\ell,\pi/\ell[$) (see Fig.~\ref{fig_1} (a) and (b)). The center of the domains containing modular variables $\bar x$ and $\bar p$ is not of further importance and was chosen for future convenience. 

The separation (\ref{eq:DefModVar}), in modular and integer parts, proved itself useful for the detection of entanglement in spatial interference patterns \cite{Clemens1,ExpModVar,mariana}. Furthermore, measurements of more general modular variables, namely periodic functions of position and momentum operators, have been used recently in proposals to test the Clauser-Horne-Shimony-Holt (CHSH)  \cite{Andreas,AsadianCHSH}, the Leggett-Garg \cite{Rabl,Saulo}  and noncontextuality inequalities \cite{CabelloContextCV,GuehneContext,AdrienContext}. Later on, in Sec.~\ref{sec:ReadoutModVar}, we provide a general framework suitable to deal with measurements of such {modular variables} based on the above introduced {modular representation}.

Further on it can be shown that the modular position and momentum operators, $\hat{\bar x}$ and $\hat{\bar p}$, commute~\cite{Aharonov}: $[\hat{\bar x}, \hat{\bar p}]=0$, which leads to the definition of the {modular basis} $\{\ket{\bar x,\bar p}|\bar x \in [-\ell/4,3\ell/4[, \bar p \in [-\pi/\ell,\pi/\ell[\}$, consisting of the common eigenstates of the these two modular operators. Consequently, in terms of the modular basis we can write $\ket{\Psi}=\int_{-\ell/4}^{3\ell/4}\int_{-\pi/\ell}^{\pi/\ell} d\bar xd\bar p\ \Psi({\bar x},{\bar p})\ \ket{\bar x,\bar p}$, with a normalized wave function $\Psi(\bar x,\bar p)$ written in the {modular representation}. The modular representation is especially convenient when dealing with periodically symmetric states, \textit{e.g.} the GKP states. As a matter of fact, using the definition of \cite{Gottesman} and the notation $\{\ket{\bar 0}, \ket{\bar 1}\}$ for the logical GKP qubits, one has simply that,  in the modular basis: $\ket{\bar x=0, \bar p=0}=\ket{\bar 0}$ and $\ket{\bar x=\ell/2, \bar p=0}=\ket{\bar 1}$, showing that the  GKP states naturally emerge from the modular basis. A formal definition of the modular basis and related expressions, as well as an example of $\Psi(\bar x,\bar p)$ will be discussed in the next section.

Formally, the modular eigenstates $\ket{\bar x,\bar p}$ can be defined as superposition of position or momentum eigenstates (distinguished by subscripts $x$ and $p$, respectively):
\begin{eqnarray}
|\bar x, \bar p\rangle &=&\sqrt{\frac{\ell}{2\pi}}\sum_{n=-\infty}^{+\infty}
e^{i\bar{p} n \ell }  |\bar x+ n \ell\rangle_x,\\
&=& \sqrt{\frac{1}{\ell}}
e^{-i\bar p \bar x}\sum_{m=-\infty}^{+\infty}
e^{-i2\pi m \bar x/\ell}
|\bar p+m 2\pi/\ell\rangle_p ,
\label{app:eq:ModVarEigenstates}
\end{eqnarray}
fulfilling the completeness and the orthogonality relation:
\begin{eqnarray} 
&&\mathbb 1=\int_{-\ell/4}^{3\ell/4}d\bar x \int_{-\pi/\ell}^{\pi/\ell} d\bar p\ket{\bar x,\bar p}\bra{\bar x,\bar p}, \\
&&\braket{\bar x,\bar p}{\bar{x}',\bar{p}'}=\delta^{(\ell)}(\bar x-\bar{x}') \delta^{(2\pi/\ell)}(\bar p-\bar{p}').
\label{app:eq:CompOrthog}
\end{eqnarray}
where $\delta^{(\ell)}(\bar x)$ and $\delta^{(2\pi/\ell)}(\bar p)$ are Dirac $\delta$ functions defined on the intervals $[-\ell/4,3\ell/4[$ and $[-\pi/\ell,\pi/\ell[$, respectively (for brevity, we will omit in the following the superscripts $\ell$ and $2\pi/\ell$). Inversely, we can define the position and momentum eigenstates in terms of the new modular eigenstates, as:
\begin{eqnarray}\label{app:eq:PosStatesModVarStates}
\ket x_x&=&\ket{\bar x+n\ell}_x= \sqrt{\frac{\ell}{2\pi}} \int_{\pi/\ell}^{\pi/\ell} \text d \bar p\ e^{-i \bar p n \ell} \ket{\bar x,\bar p}, \\
\ket p_p&=&\ket{\bar p+m\frac{2\pi}{\ell}}_p= \sqrt{\frac{1}{\ell}} \int_{-\ell/4}^{3\ell/4}d\bar x\ e^{i\bar x \bar p }e^{i2\pi m\bar x /\ell} \ket{\bar x,\bar p}.  \nonumber \\ \label{app:eq:MomStatesModVarStates}
\end{eqnarray}
Hence, an arbitrary state in position representation $\ket{\Psi}= \int \text dx\ \Psi(x) \ket{x}$ transformed to the modular representation, reads:
\begin{eqnarray}
\ket\Psi=\int_{-\ell/4}^{3\ell/4}\int_{-\pi/\ell}^{\pi/\ell} d\bar xd\bar p \Psi(\bar x,\bar p)
\ket{\bar x,\bar p}, 
\label{app:eq:ModVarState}
\end{eqnarray}
where 
\begin{align}
\Psi(\bar x,\bar p)=\sqrt{\frac{\ell}{2\pi}}\sum_{n=-\infty}^{\infty} \Psi(n\ell+\bar x) e^{- i n \bar p \ell},
\label{eq:ModularWaveFct}
\end{align}
is called {modular wave function} of $\ket\Psi$. The same representation was introduced by J. Zak in 1967 under the term $k,q$-representation \cite{Zak1967}.

\subsection{Examples of modular wave functions}
The modular variables representation is particularly well suited for wave functions that obey a certain periodicity in position or momentum space.  For example, the state $\ket{\psi_{c}}=\sum_{n=-\infty}^{+\infty} \ket{L n}_x$, representing a comb of $\delta$ functions with distance $L$ in position space, becomes in the modular representation $\ket{0,0}$, namely a single $\delta$ peak at the origin, if we set $\ell=L$. This state, together with $\ket{\ell/2,0}$, are examples of the logical qubit state introduced in the GKP paper \cite{Gottesman}.

Instead, a more physical state can be obtained if we replace the $\delta$ comb by a comb of finitely squeezed Gaussian spikes with width $\Delta$ and a Gaussian envelope with width $1/\kappa$ (see Fig.~\ref{fig_2}(b)). The wave function of such a state in position representation, reads:
\begin{align}
\Psi_{G,c}(x)=\frac{N}{(\pi \Delta^2)^{\frac{1}{4}}} e^{-(x \kappa)^2/2} \sum_{n=-\infty}^{\infty} e^{-(x-n L)^2/2\Delta^2},
\label{app:eq:approxGKP}
\end{align}
with a normalization factor $N$. In the limit $\Delta/L \ll1$ and $\kappa L \ll 1$, of a large envelope and sufficiently thin spikes, respectively, the latter can be approximated by $N\approx\sqrt{L \kappa/\sqrt\pi}$. Then, transforming Eq. (\ref{app:eq:approxGKP}) to the modular representation with the help of Eq. (\ref{app:eq:ModVarEigenstates}), yields:
\begin{align}
\Psi_{G,c}(\bar x,\bar p)=T(\bar x) C(\bar p), 
\label{app:eq:approxGKPmodular}
\end{align}
where
\begin{align}
T(\bar x)&=\frac{1}{(\pi\Delta^2)^{\frac{1}{4}}} \sum_{n}e^{-(\bar x-n\ell)^2/2\Delta^2},\\
C(\bar p)&=\frac{1}{(\pi\kappa^2)^{\frac{1}{4}}} \sum_{m}e^{-(\bar p-m2\pi/\ell)^2/2\kappa^2},
\label{app:eq:GKPmodular}
\end{align}
and $\ell=L$. To obtain the above result we used that according to the Poisson sum formula we have $\sqrt a \sum_{m}e^{-\pi a(m-b)^2}= \sum_n e^{2\pi in b} e^{-\pi n^2/a}$, and that in the limit of large Gaussian envelopes we can approximate $e^{-x \kappa^2/2}\approx e^{- (nL\kappa)^2/2}$. 
A possible experimental platform allowing for the production of such periodic states is given by the transverse degrees of freedom of single photons, as illustrated in Fig.~\ref{fig_1}(c).

\section{Quantum information processing framework} \label{sec:QIPframework}
\subsection{Logical states}
Labeling quantum states using bounded continuous variables enables the definition of two disjoint sets of equal size for each one of the variables. Such a splitting can be done in infinitely many ways, and in order to illustrate the principles of our ideas we discuss in detail the splitting of the domain of the variable $\bar x$ into two subintervals $[-\ell/4,\ell/4[$ and $[\ell/4,3\ell/4[$. As a consequence, we obtain a continuum of two-level systems spanned by the states $\{\ket{{\bar x},{\bar p}},\ket{{\bar x}+\ell/2,{\bar p}}\}$ in terms of which we can express a general state $\ket\Psi$ as:
\begin{align}
\ket{\Psi}=\int_{-\ell/4}^{\ell/4}d\bar x \int_{-\pi/\ell}^{\pi/\ell} d\bar p f(\bar x,\bar p) \ket{\tilde{\Psi}(\bar x,\bar p)},
\label{eq:GeneralState}
\end{align}
where
\begin{align}
\ket{{\tilde\Psi}(\bar x,\bar p)}=&\cos{\left(\frac{\theta(\bar x,\bar p)}{2}\right)} \ket{\bar x,\bar p} \nonumber \\
& + \sin{\left(\frac{\theta(\bar x,\bar p)}{2}\right)} e^{i\phi(\bar x,\bar p)} \ket{\bar x+\ell/2,\bar p}.
\label{eqn:continuousqubit}
\end{align} 
with a complex function $f(\bar x,\bar p)$ such that  $\int_{-\ell/4}^{\ell/4}d\bar x \int_{-\pi/\ell}^{\pi/\ell} d\bar p |f(\bar x,\bar p)|^2 =1$ and two real functions, $\theta(\bar x,\bar p)$ and $\phi(\bar x,\bar p)$, defined on $[-\ell/4,\ell/4[\times[-\pi/\ell,\pi/\ell[$. The mathematical expressions allowing to switch back and forth between the position and modular representation can be found in Appendix~\ref{app:LogicalQubits}. Equation~(\ref{eq:GeneralState}) can be seen as  a weighted continuous superposition of  pure qubit states $\ket{\tilde{\Psi}(\bar x,\bar p)}$ for each subspace with fixed $\bar x$ and $\bar p$. We stress that, so far, no approximation has been made, and state (\ref{eq:GeneralState}) is simply an alternative way of writing an arbitrary state expressed in a continuous basis. Note that the choice of $\ell$ is also arbitrary, and modifying it for a given state modifies the definition of the functions appearing in (\ref{eq:GeneralState}) and (\ref{eqn:continuousqubit}). 

In the following, in order to encode discrete quantum information in CV states, we assume that $\theta(\bar x,\bar p)=\theta$ and $\phi(\bar x,\bar p)=\phi$ are constant functions such that Eq.~(\ref{eq:GeneralState}) becomes $\ket\Psi=\cos{\left(\theta/2\right)} \ket{0_L} + \sin{\left(\theta/2\right)} e^{i\phi} \ket{1_L}$ with  logical qubit states, defined as:
\begin{align}
\ket{0_L}&=\int_{-\ell/4}^{\ell/4}d\bar x \int_{-\pi/\ell}^{\pi/\ell} d\bar p f(\bar x,\bar p) \ket{\bar x,\bar p}, \label{eq:Logical0} \\
\ket{1_L}&=\int_{-\ell/4}^{\ell/4}d\bar x \int_{-\pi/\ell}^{\pi/\ell} d\bar p f(\bar x,\bar p) \ket{\bar x+\ell/2,\bar p},\label{eq:Logical1}
\end{align}
The logical qubit states (\ref{eq:Logical0}) and (\ref{eq:Logical1}) reflect a dichotomization of the Hilbert space with respect to the modular position $\bar x$. The exact choice of $f(\bar x,\bar p)$ is arbitrary as long as it emerges from a  properly defined modular wave function $\Psi(\bar x,\bar p)$ (see Appendix~\ref{app:LogicalQubits}). 
In Fig.~\ref{fig_2}(a) we plot the modulo square of these states in the case where $f(\bar x,\bar p)$ is given by a two dimensional Gaussian function centered at the origin with standard deviations $\Delta$ and $\kappa$ in the modular position and momentum variables, respectively. In position space the same states are represented by two shifted combs of Gaussian spikes with width $\Delta$ and Gaussian envelope of width $\kappa^{-1}$ (see Fig.~\ref{fig_2}(b)). This example corresponds to the well known non-perfect GKP states introduced in \cite{Gottesman}. 

We now discuss the manipulation of the introduced logical states through appropriate unitary operations and their analogy to ordinary Pauli  matrices.

\begin{figure}[t]
\includegraphics[scale=0.475]{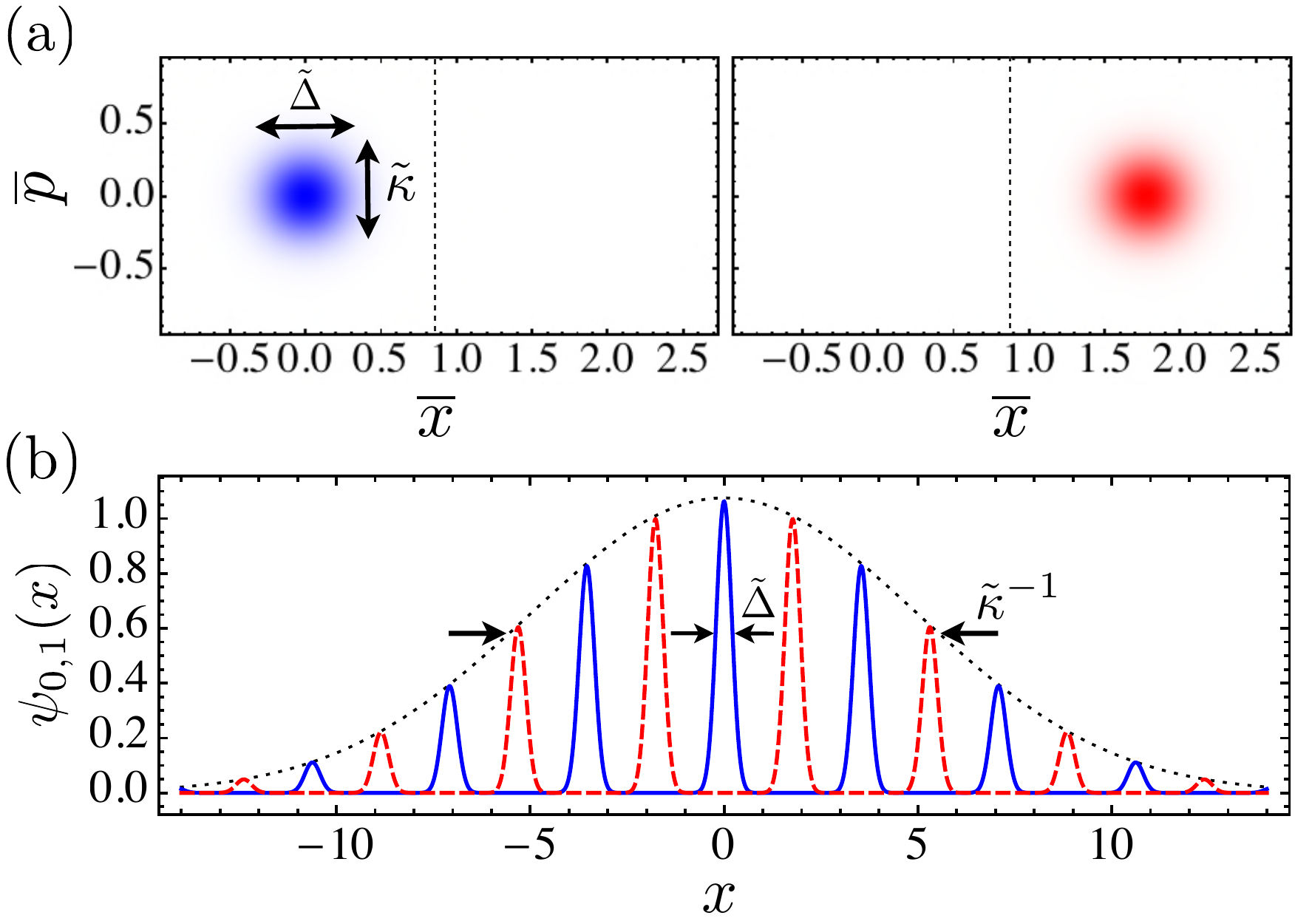}
\caption{(color online) (a) Density plots of the probability distributions of the logical states $\ket{0_L}$ and $\ket{1_L}$, respectively, in the modular representation (see Eq. (\ref{eq:GeneralState})) with $f(\bar x,\bar p)$, $\theta(\bar x,\bar p)$ and $\phi(\bar x,\bar p)$ chosen as explained in the text and $\ell=2\sqrt\pi$.  The full widths at half maximum (FWHM) of the distributions are indicated by $\tilde\Delta=2\sqrt{2\ln{2}}\Delta$ and $\tilde\kappa=2\sqrt{2\ln{2}}\kappa$. (b) Plot of the wave function $\psi_{0,1}(x)$ of the same two logical states in the position representation. In momentum space the same functions with switched roles of $\tilde\Delta$ and $\tilde\kappa$ represent the logical states $\ket{\pm_L}=\frac{1}{\sqrt 2}(\ket{0_L} \pm \ket{1_L})$.  }
\label{fig_2}
\end{figure}

\subsection{Logical operations}
In this section, we introduce some single qubit logical operations acting on the qubit structure (\ref{app:eq:ContQubitRep2}) that is naturally embedded in every state $\ket\Psi$. We start by expressing the single mode phase space displacement operator $\hat D(\Delta x,\Delta p)=\exp(i \Delta p \hat x-i \Delta x \hat p)=\exp(i \Delta p \hat x)\exp(-i \Delta x \hat p)\exp(-i \Delta x\Delta p/2)$ in the modular representation. To do so, we calculate first its action on a modular eigenstate (\ref{app:eq:ModVarEigenstates}), yielding:
\begin{eqnarray}
\hat D(\Delta x,\Delta p)&=&e^{-i \frac{\Delta x\Delta p}{2}}\int_{-\ell/4}^{3\ell/4}d\bar x \int_{-\pi/\ell}^{\pi/\ell} d\bar p e^{i(\bar p+\Delta p)(\overline{\bar x+\Delta x})}  \nonumber \\
&\times& e^{-i\bar p (\bar x+\Delta x)} \ket{\overline{\bar x+\Delta x},\overline{\bar p+\Delta p} }  \bra{\bar x,\bar p}. 
\label{app:eq:DispOp}
\end{eqnarray}
Over-lined expressions denote the corresponding modular parts in position or momentum, respectively. 
Equation (\ref{app:eq:DispOp}) shows that a phase space displacement by $(\Delta x,\Delta p)$ leads to a displacement of the corresponding modular position and momentum accompanied by the generation of additional phase factors. The latter encode information about the change of the discrete position and momentum values, $\ell\hat N$ and $2\pi/\ell \hat M$, induced by the displacement.

\subsubsection{Pauli and Weyl-Heisenberg operators} 
Now, by setting the displacements in Eq.~(\ref{app:eq:DispOp}) equal $(\Delta x=0$, $\Delta p=2\pi/\ell)$ and $(\Delta x=\ell/2$, $\Delta p=0)$, respectively, and by splitting the $\bar x$-integration we reveal the following two operators:
\begin{eqnarray}
\hat{ Z}=e^{2 \pi  i\hat x /\ell}=\int_{-\ell/4}^{\ell/4}d\bar x \int_{-\pi/\ell}^{\pi/\ell} d\bar p \ e^{2\pi i \bar x /\ell} \hat\sigma_z(\bar x,\bar p),
\label{eq:SigmaZ}\\
\hat{ X}=e^{-i\hat p \ell/2}=\int_{-\ell/4}^{\ell/4}d\bar x \int_{-\pi/\ell}^{\pi/\ell} d\bar p \ e^{-i\bar p \ell/2} \hat\sigma_x(\bar x,\bar p),
\label{eq:SigmaX}
\end{eqnarray}
where 
\begin{eqnarray}
\hat\sigma_x(\bar x,\bar p)&=& \ket{|{\bar x,\bar p} }  \bra{{\bar x+\ell/2,\bar p}|}+ \ket{|{\bar x,\bar p}}  \bra{{\bar x+\ell/2,\bar p}|}, \label{eq:sigz} \\
\hat\sigma_z(\bar x,\bar p)&=& \ket{|{\bar x,\bar p} }  \bra{{\bar x,\bar p}|} - \ket{|{\bar x+\ell/2,\bar p} }  \bra{{\bar x+\ell/2,\bar p}|}, \label{eq:sigx}
\end{eqnarray}
with $\{\ket{|\bar x,\bar p}, \ket{|\bar x+\ell/2 ,\bar p}\}=\{e^{-i\bar p \ell/4}\ket{\bar x,\bar p}, e^{i\bar p \ell/4}\ket{\bar x+\ell/2 ,\bar p}\}$. We thus see that the Weyl-Heisenberg operators act as ordinary Pauli operators ($\hat\sigma_{z}$ and $\hat\sigma_{x}$) on each of the qubit subspaces defined by $\{\ket{|\bar x,\bar p}, \ket{|\bar x+\ell/2 ,\bar p}\}$. The analog of the third Pauli operator $\hat\sigma_y$ can be obtained from the product of the former two $\hat Y=i\hat X^\dagger \hat Z^\dagger$, yielding:
\begin{align}
\hat Y=\int_{-\ell/4}^{\ell/4} d\bar x  \int_{-\pi/\ell}^{\pi/\ell} d\bar p \  e^{i\bar p \ell/2-2\pi i \bar x/\ell} \hat\sigma_y(\bar x,\bar p),
\label{eq:SigmaY}
\end{align}
with
\begin{align}
\hat\sigma_y(\bar x,\bar p)=i( \ket{|{\bar x,\bar p} }  \bra{{\bar x+\ell/2,\bar p}|}- \ket{|{\bar x,\bar p}}  \bra{{\bar x+\ell/2,\bar p}|}). \label{eq:SigmaxpY}
\end{align}
On the other hand, using the commutation rules for the phase space displacement operator, we find that $\hat X^\dagger \hat Z^\dagger=-i \hat D^\dagger(\ell/2,2\pi/\ell)$, yielding $\hat Y=\hat D^\dagger(\ell/2,2\pi/\ell)$. Thus, as illustrated in Fig.~\ref{fig_1}(b), the three displacements implementing the logical Pauli operations (\ref{eq:SigmaZ}), (\ref{eq:SigmaX}) and (\ref{eq:SigmaY}), form a triangular in phase space that encloses an area of $\pi/2$. The latter is closely related to the fact that the anti-commutators between the displacements (\ref{eq:SigmaX}), (\ref{eq:SigmaZ}) and (\ref{eq:SigmaY}), vanish \citep{GuehneContext}, yielding the anti-commutation relations of our logical Pauli operators:
\begin{align}
\{\hat Z,\hat X\}=\{\hat Z,\hat Y\}=\{\hat X,\hat Y\}=0,
\label{eq:AntiCommutators}
\end{align}
as expected form the algebra of Pauli matrices.

However, despite the similarities, the logical operations defined above are not completely equivalent to a Pauli algebra in the general case. This becomes apparent from their commutation relations which are found to be $[\hat X,\hat Y]=2i\hat Z^\dagger$, $[\hat Z,\hat X]=2i\hat Y^\dagger$ and $[\hat Y,\hat Z]=2i\hat X^\dagger$. They resemble those of the Pauli matrices, but since the operators $\hat X$, $\hat Y$ and $\hat Z$ are not hermitian, they deviate in the fact that the commutator between each of them yields the adjoint of the third one. We further note that, in the subspace spanned by the GPK states $\ket{\bar 0}$ and $\ket{\bar 1}$, the operators $\hat X$, $\hat Y$ and $ \hat Z$ are Hermitian ($\hat X=\hat X^\dagger$, $\hat Y=\hat Y^\dagger$ and $\hat Z=\hat Z^\dagger$). Hence, in this subspace, the above commutation relations become equal to those of a real Pauli algebra. 

The fact that we are dealing with a nonperfect Pauli algebra has some consequences. For instance, if we calculate the square of one of the logical Pauli operators we get:
\begin{equation}
\hat Z^2=e^{4 \pi  i\hat x /\ell}=\int_{-\ell/4}^{\ell/4}d\bar x \int_{-\pi/\ell}^{\pi/\ell} d\bar p \ e^{4\pi i \bar x /\ell}  \mathbb{1} (\bar x,\bar p),
\label{eq:logicalZsquared}
\end{equation}
 with $\mathbb{1} (\bar x,\bar p)=\ket{|{\bar x,\bar p} }  \bra{{\bar x,\bar p}|} + \ket{|{\bar x+\ell/2,\bar p} }  \bra{{\bar x+\ell/2,\bar p}|}$
, which differs from an identity through the appearance of an $\bar x$ dependent phase factor under the integral.  Similarly, such phase factors also appear when manipulating the states $\ket{0_L}$ and $ \ket{1_L}$ with one of the logical operations (\ref{eq:SigmaZ}), (\ref{eq:SigmaX}) or (\ref{eq:SigmaY}). We will see later on that these phase factors become irrelevant if one considers only protocols involving a specific class of modular variables as readout observables. In light of the definition of these readout observables, which will be given in Sec.~\ref{sec:ReadoutModVar}, we will also introduce appropriate rotation operators allowing to perform measurements according to different mutually unbiased bases of the logical space (see Sec.~\ref{sec:ModularRotations}). Consequently, the logical states and Pauli operations, together with the modular readout observables and the corresponding rotations (see Secs.~\ref{sec:ReadoutModVar} and \ref{sec:ModularRotations}), establish a solid framework to handle CV quantum information from a quantum measurement point of view. 

\subsubsection{Clifford gates:} A qubit phase gate $\hat P$ can be realized using the shear operation $e^{i \hat x^2/(2 d^2)}$ with $d=\ell/(2\sqrt\pi)$. It transforms the logical Pauli operators, (\ref{eq:SigmaZ}) and (\ref{eq:SigmaX}), as:
\begin{eqnarray}
\hat{X}\rightarrow \int_{-\ell/4}^{\ell/4}d\bar x \int_{-\pi/\ell}^{\pi/\ell} d\bar p e^{2\pi i \bar x/\ell-i\bar p\ell/2} \hat\sigma_y(\bar x,\bar p)=i\hat Z \hat X,
\label{app:eq:SigmaY}
\end{eqnarray}
and $\hat Z\rightarrow \hat Z$. In this case, the shear implements a rotation of $\hat X$ around the $z$-axis of the Bloch sphere. Further on, the Hadamard gate $\hat H$ can be directly realised using a rescaled Fourier transform $\hat{\cal F}=e^{i\frac{\pi}{4}(\hat x^2/d^2+\hat p^2 d^2)}$, with $d$ chosen as above, which transforms the logical Pauli operators as $\hat X\rightarrow \hat Z$ and $\hat Z\rightarrow \hat X^{-1}$. In combination with the above defined logical phase-gate, we can define the Fourier transformed shear $e^{i \hat p^2d^2/2}$ which then implements a $\pi/2$-rotation of $\hat Z$ around the $x$-axis, namely $\hat Z\rightarrow \hat Y$ and $\hat X\rightarrow -\hat Y$. 
Finally, the two-qubit Clifford generator $\hat C_{\text{NOT}}$ can be realized by the two-mode Gaussian unitary $e^{-i \hat x_a\otimes\hat p_b}$ which implements the operations $\hat X^a\otimes \hat X^b \rightarrow \hat X^a\otimes  \hat X^{b-a}$ and $\hat Z^a\otimes \hat Z^b \rightarrow \hat X^{a+b}\otimes  \hat Z^{b}$, with $a,b=0,1$. Note, that the logical controlled-phase gate $\hat C_{Z}$ follows from $\hat C_{\text{NOT}}$ by an additional application of $\hat{\cal F}$ on the second mode.

Note that these logical operations implement the desired Clifford group operation  only when acting on perfect GKP logical states $\ket{0,0}$ and $\ket{\ell/2,0}$. Therefore, the finite squeezing of the logical states $\ket{0_L}$ and $\ket{1_L}$  leads to a faulty implementation of the above defined logical Clifford operations which manifests itself by a washing out of the signal \cite{MBQC}. In order to  circumvent such errors one can apply GKP error correction to the encoded states \cite{Gottesman,errorboundsGKP,MenicucciFault}, which keeps the squeezing on a tolerable level. This problem does not occur if we manipulate our logical qubits with the rotation operator defined in Sec.~\ref{sec:ModularRotations}. An experimental implementation of these rotations using the transverse degrees of freedom of single photons will be discussed later on in Sec.~\ref{sec:ExpProposal}.

We will move on now and show how measurements of suitably defined modular variables allow to retrieve binary quantum information that is encoded in terms of our modular logical states.

\subsection{State readout with modular variables} \label{sec:ReadoutModVar}
We start by defining the observables $\hat \Gamma_{\beta}$, which are the analogous to the Pauli matrices  from the point of view of a measurement: 
\begin{eqnarray}
\hat{\Gamma}_\beta=\int_{-\ell/4}^{\ell/4}d\bar x \int_{-\pi/\ell}^{\pi/\ell} d\bar p\  \zeta_\beta (\bar x, \bar p) \hat\sigma_\beta(\bar x,\bar p),
\label{eq:Gammas}
\end{eqnarray}
with $\beta=x,y,z$, $\zeta_\beta(\bar x,\bar p)$ a real and bounded function and $\hat\sigma_\beta(\bar x,\bar p)$ defined as in Eqs.~(\ref{eq:sigz}), (\ref{eq:sigx}) and (\ref{eq:SigmaxpY}). 
As we show in Appendix~\ref{app:GammaProp}, the sum over $\beta$ of the modulo squares of the expectation value of these observables is bounded by $( \max_{\bar x,\bar p,\beta}|\zeta_\beta{(\bar x,\bar p)}|)^2$. Moreover, restricted to our logical space, defined by $\ket{0_L}$ and $\ket{1_L}$, these expectation values can be expressed as 
\begin{align}
\moy{\hat{\Gamma}_\beta}= K_\beta \moy{\hat\sigma_\beta}, 
\end{align}
where
\begin{align}
K_\beta=\int_{-\ell/4}^{\ell/4}d\bar x \int_{-\pi/\ell}^{\pi/\ell} d\bar p\  \zeta_\beta (\bar x, \bar p) |f(\bar x,\bar p)|^2,
\end{align}
 and $\moy{{\boldsymbol\sigma}}=(\sin{\theta}\cos{\phi},\sin{\theta}\sin{\phi},\cos{\theta})$. Hence, we find that the expectation values of the observables (\ref{eq:Gammas}) are proportional to the Bloch vector of the encoded qubit states, whereas the proportionality factors are determined by the overlap of $|f(\bar x,\bar p)|^2$ with $\zeta_\beta(\bar x,\bar p)$. We can also define spatial and temporal correlators among observables of the kind of Eq.~(\ref{eq:Gammas}), which have already been proven to be useful in the context of testing quantum mechanical properties in CV in \cite{Andreas,AdrienContext,Pierre,Rabl,Rabl1,GuehneContext}. All these works involve measurements of particularly chosen modular variables that can be expressed in the form of Eq.~(\ref{eq:Gammas}).

The form of the operators (\ref{eq:Gammas}) is chosen so as to be operationally analogous to the logical Pauli operations, defined in Eqs.~(\ref{eq:SigmaZ}), (\ref{eq:SigmaX}) and (\ref{eq:SigmaY}). Interestingly, unwanted phase factors, appearing when manipulating the states $\ket{0_L}$ and $ \ket{1_L}$ with some logical operation, disappear. For instance, if we consider the operator $\hat Z^2$ (see Eq.~(\ref{eq:logicalZsquared})) 
and apply it to an arbitrary state of the logical space $\ket{\Psi}=\cos{\theta}\ket{0_L}+\sin{\theta} e^{i\phi}\ket{1_L}$, we obtain $\ket{\Psi'}=\hat Z^2 \ket{\Psi}$, where $\ket{\Psi'}$ differs from $\ket\Psi$ by a modular position (momentum) dependent phase factor, but we have $\langle \hat \Gamma_\beta \rangle_{\psi}=\langle \hat \Gamma_\beta \rangle_{\psi'}$, for all $\beta=x,y,z$. Therefore, for implementations of protocols involving measurements of the expectation values $\langle \hat \Gamma_\beta \rangle_{\ket\psi}$, the $\hat Z^2$ operator acts as the identity. Similarly, phase factors that appear due to the application of the logical Pauli operations (\ref{eq:SigmaZ}), (\ref{eq:SigmaX}) and (\ref{eq:SigmaY}) to a logical state $\ket\Psi$,  are invisible to measurements of the expectation values of (\ref{eq:Gammas}). Consequently, this allows one to establish a solid framework for handling discrete quantum information encoded in the CV logical states $\ket{0_L}$ and $ \ket{1_L}$. 

In Appendix~\ref{app:Conditions}, we discuss the conditions a general phase space observable $F(\hat x,\hat p)$, where $F$ is a real-valued function, has to fulfill such that it can be written in the form (\ref{eq:Gammas}). In general, we find that the observables (\ref{eq:Gammas}) are given by periodic phase space observables, meaning observables that can be expressed as:
\begin{align} 
F(\hat x,\hat p)&= \sum_{n=-\infty}^\infty  \sum_{m=-\infty}^\infty  d_{n,m} e^{2\pi i n \hat x/L-iL' m \hat p},
\label{eq:DoubleFourier}
\end{align}
where $d_{n,m}$ are complex Fourier coefficients obeying the normalization condition $ \sum_{n=-\infty}^\infty  \sum_{m=-\infty}^\infty  |d_{n,m}|^2=1$, and $L$ and $L'$ denote to the periodicities in position and momentum of the corresponding phase space representation of $F(\hat x,\hat p)$. All three observables (\ref{eq:Gammas}) can be expressed as (\ref{eq:DoubleFourier}) with different choices of $d_{n,m}$, $L$ and $L'$ (see Appendix~\ref{app:Conditions}). 

Examples of such observables are, for instance, $\text{Re}(\hat X)$, $\text{Re}(\hat Y)$ and $\text{Re}(\hat Z)$, where $\zeta_x(\bar x,\bar p)=\cos{(\bar p \ell/2)}$, $\zeta_y(\bar x,\bar p)=\cos{(2\pi \bar x/\ell-\bar p\ell/2)}$ and $\zeta_z(\bar x,\bar p)=\cos{(2\pi  \bar x/\ell)}$. We note that the general definition of the observables (\ref{eq:Gammas})  leads only in the case $\zeta_\beta(\bar x,\bar p)=1$, for all $\beta$, to a real set of Pauli operators. However, if one wants to favour the experimental accessibility of such observables via positive operator valued measurements (see Sec.~\ref{sec:MeasurementPOVM} and Ref.~\cite{Horodecki}), it is desirable to keep the freedom of choice of the functions $\zeta_\beta(\bar x,\bar p)$. This leads, in the general case where $\zeta_\beta(\bar x,\bar p)$ is given by a continuous function, to an operator with a continuous spectrum. 

\subsection{Arbitrary rotations}\label{sec:ModularRotations}
Once we have created the possibility of retrieving quantum information through measurement of binary observables defined in CV, one can complete the set of qubit like operations by defining arbitrary rotations on the encoded subspaces. However, since the operators $\hat X$, $\hat Y$ and $\hat Z$ are unitary but not hermitian, arbitrary rotations cannot be generated by simply exponentiating them with the proper multiplicative factors, as is true for the Pauli matrices \cite{Nielsen}. This can be done only if we consider the operators (\ref{eq:Gammas}) in the special case, $\zeta_\beta(\bar x,\bar p)=1$, for all $\bar x$, $\bar p$ and $\beta$ (in the following denoted by $\hat{\Gamma}^1_\beta$), which then can be used to define  
\begin{eqnarray}
e^{i\frac{\phi}{2}(\hat{{\boldsymbol\Gamma}}^1\cdot \textbf n)}=\cos{\left(\frac{\phi}{2}\right)}  \mathbb 1 + i \sin{\left(\frac{\phi}{2}\right)} (\hat{{\boldsymbol\Gamma}}^1\cdot \textbf n),
\label{eq:Rotations}
\end{eqnarray}
where $\hat{\boldsymbol\Gamma}^1=(\hat\Gamma_x^1,\hat\Gamma_y^1,\hat\Gamma_z^1)$ and $\textbf n=(n_x,n_y,n_z)$ indicates the axis of rotation. Equation (\ref{eq:Rotations}) allows to perform rotations of the general observables (\ref{eq:Gammas}) without changing the function $\zeta_\beta(\bar x,\bar p)$ and thus to implement measurements in different mutually unbiased bases of the corresponding logical space. Note that, in contrary to logical operations introduced in \cite{Gottesman}, the operators (\ref{eq:Rotations}) perform well not only on the subspace spanned by perfect GKP states but also on the space spanned by the more general logical states  $\ket{0_L}$ and $\ket{1_L}$.  A proposal of an experimental implementation of these rotation operations using the spatial distribution of single photons is discussed in Sec.~\ref{app:ImplementationLogOps}. 

\section{Proposal of experimental implementations} \label{sec:ExpProposal}
In the following, we assume that the coordinates $\hat x$ and $\hat p$ refer to the transverse position and momentum of a single photon. These variables are related to the object or source plane (position plane) and the Fourier plane (momentum plane) of a single-photon field. If we remain in the paraxial approximation, $\sin{\theta}\approx \theta$ (see Fig.~\ref{fig_1}(c)), the wave function of this field can be seen as the wave function of a single point particle, here being the photon. We restrict ourselves to the one dimensional case because the Hilbert space associated to the two dimensional spatial photon field is a tensor product of the Hilbert spaces associated with the two orthogonal transverse directions of the photon \cite{Tasca}. A general quantum state of the transverse momentum (or position) of the photon can be written in the {modular basis}, as shown in (\ref{app:eq:ContQubitRep1}).

\subsection{Creation of states with periodic wave function}\label{sec:CreationWaveFkt}
\subsubsection{Single photons}
\begin{figure}[t]
\includegraphics[width=0.5\textwidth]{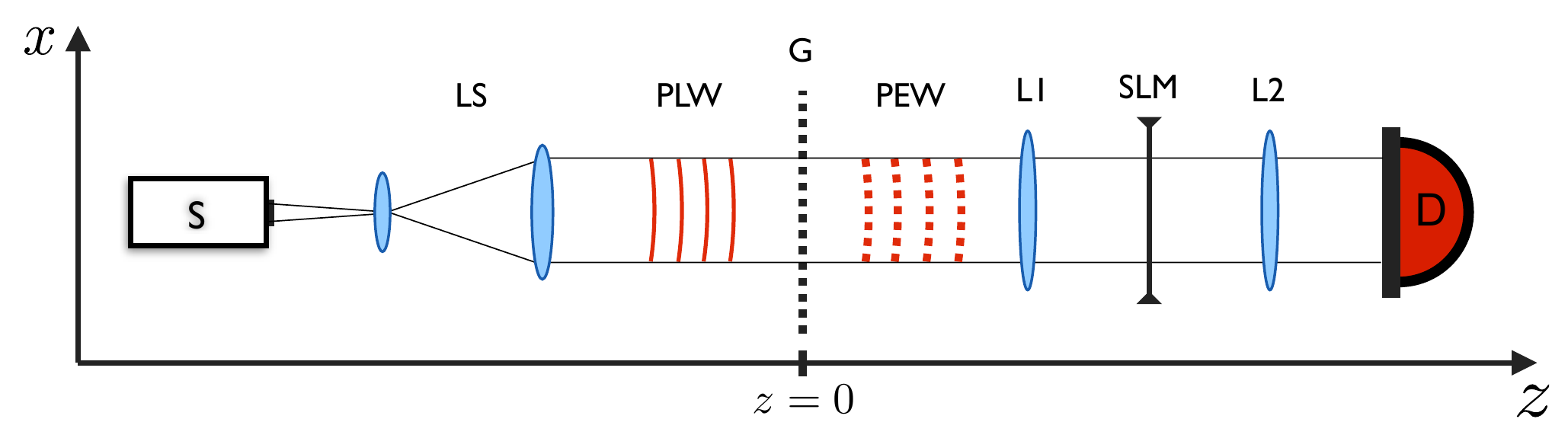}
\caption{(color online) Scheme showing the production, processing and detection of the transverse degrees of single photons. The photons are produced in a source (S) and then send through a lens systems (LS) in order to prepare them in approximate plane waves. Further on, a periodic refraction grating prepares the photon's transverse wave function in a periodic state. The photons can be manipulated using a spatial light modulator (SLM) implementing the unitary operation $\hat U_{\text{SLM}}$, or optionally $\hat{\cal F}\hat U_{\text{SLM}}\hat{\cal F}^\dagger$ if placing lenses (L1,L2) before and after the SLM. Finally, the photons are measured with a spatially resolving detector (D).  }
\label{fig_3} 
\end{figure}

One major advantage in using the transverse degrees of freedom of single photons is that we can very efficiently produce states with a periodic wave function, as those presented in Fig.~\ref{fig_2}. To do so, we simply pass the photons through a periodic diffraction grating, as indicated in Fig.~\ref{fig_3}. If the photon, which is impinging on the grating, has a Gaussian transverse wave function $f_G(x)\propto e^{-(x \kappa)^2/2}$, with width $\kappa^{-1}$, and the transmission function of the grating is given by $\sum_m a_m e^{i m x 2\pi/L}$, where $a_m=\exp{[-\frac{1}{2}m^2 (2\pi \Delta/L)^2]}$ with slit width $\Delta$ and distance $L$, the resulting wave function of the diffracted photons has the form~(\ref{app:eq:approxGKP}). Hence, by by adjusting the slit widths and distances of the grating, we can produce the logical qubit states $\ket 0_L$ and $\ket 1_L$. In Fig.~\ref{fig_3}, the photons are sent through a lens system before the grating in order to prepare them in approximate plane waves. This allows us additionally to adjust the width $\kappa^{-1}$ of the Gaussian envelope of the wave function (\ref{app:eq:approxGKP}), which corresponds to the quality of the prepared plane waves. Experimentally gratings are often realized using spatial light modulators (SLMs) (see Fig.~\ref{fig_1}(c)). 

Note that the free propagation of the diffracted photons will lead to a blurring in the photon's transverse wave function. This effect is mainly due to the finite envelope of the photons transverse wave function and can be minimized by improving the quality of the prepared plane waves before the grating. The dependency of the blurring on the number of irradiated slits was studied in \cite{Talbot} by calculating the fidelity of the initially prepared wave function with its revivals after multiples of a specific propagation distance, referred to as Talbot distance. It was shown that, for currently available diffraction gratings, a fidelity higher than $0.9$ can be maintained for a propagation distance of about 10 times the Talbot distance. For larger propagation distances one could possibly improve the fidelity using the earlier mentioned GKP error correction procedure to the transverse field of the photons.

\subsubsection{Entangled photon pairs}
Once it is known how to prepare single photons in the corresponding logical states $\ket{0_L}$ and $\ket{1_L}$, we can try to entangle a pair of photons and further use it for the realization of quantum information protocols. One possibility to do so is by producing polarization entangled states in a type-2 parametric down-conversion process and subsequently swapping the polarization entanglement to the spatial distributions of the two photons, as suggested in \cite{Andreas}. Therefore, the photons are sent through a Mach-Zehnder interferometer made up of polarization dependent beam-splitters, half- and quater-wave plates and diffraction gratings. The output states of the interferometers will yield to 50\% the desired spatially entangled photons, while the other half of the emerging photons has to be discarded. Details of the procedure can be found in \cite{Andreas}. 

\subsection{Logical operations realized by linear optical elements}\label{app:ImplementationLogOps}

\subsubsection{Spatial light modulator}
All our following discussion about experimental implementations of single qubit rotations or measurements of readout observables rely on an optical element, called spatial light modulator (SLM). A SLM consists usually of some transparent or reflective  screen that is divided into a certain number of pixels whose diffraction index can be adjusted individually.  In this way, one can impose spatial phase or amplitude modulations on a light beam that is transmitted or reflected by the SLM. In particular, it is possible to implement operations of the form $\hat U_{\text{SLM}}=e^{i h(\hat x)}$, where $h(x)$ is an arbitrary user-defined function. Similarly, one can implement phase modulations in momentum space by combining a SLM with the Fourier transform  $\hat{\cal F}=e^{i\frac{\pi}{4}(\hat x^2/d^2+\hat p^2 d^2)}$, which itself is realized optically with lenses, where $d$ is related to the focal length $f$ of the lens and the wave length $\lambda$ through the relation $d=\sqrt{f\lambda/2\pi}$ (see also Fig.~\ref{fig_3}) \cite{Tasca}. 

\subsubsection{Single qubit rotations}
The above discussion leads us to a linear optical implementation of the logical rotation operator 
\begin{eqnarray}
e^{i\frac{\phi}{2}(\hat{{\boldsymbol\Gamma}}^1\cdot \textbf n)}=\cos{\left(\frac{\phi}{2}\right)}  \mathbb 1 + i \sin{\left(\frac{\phi}{2}\right)} (\hat{{\boldsymbol\Gamma}}^1\cdot \textbf n),
\label{app:eq:Rotations}
\end{eqnarray}
where $\hat{\boldsymbol\Gamma}^1=(\hat\Gamma_x^1,\hat\Gamma_y^1,\hat\Gamma_z^1)$ and $\textbf n=(n_x,n_y,n_z)$ defines the axis of rotation. We focus on rotation around the two main axes corresponding to the operators $\hat\Gamma_z$ and $\hat\Gamma_x$, which by composition allow to implement any desired single qubit rotation. Therefore, we remind the reader that
\begin{align}
\hat\Gamma^1_ z&=\int_{-\ell/4}^{\ell/4}d\bar x \int_{-\pi/\ell}^{\pi/\ell} d\bar p\ \hat\sigma_z(\bar x,\bar p) \nonumber \\
&=\int_{-\ell/4}^{3\ell/4}d\bar x \int_{-\pi/\ell}^{\pi/\ell} d\bar p s_z(\bar x) \ket{\bar x,\bar p}\bra{\bar x,\bar p},
\label{app:eq:Gamma1z}
\end{align}
with the step function $s_z(\bar x)$ that takes the value $+1$ if $\bar x\in[-\ell/4,\ell/4[$ and $-1$ if $\bar x\in[\ell/4,3\ell/4[$. By means of the discussion in the Appendix~\ref{app:Conditions} we know that Eq. (\ref{app:eq:Gamma1z}) reads in the position representation as follows:
\begin{align}
\hat\Gamma^1_ z&=\int_{-\infty}^\infty dx  s_z(x) \ket{x}_x\bra{x}_x,
\label{app:eq:GammaZPos}
\end{align} 
where $s(x)$ is a $\ell$-periodic rectangular function taking the value $+1$ if $x\in [-(2n)\ell/4,(2n)\ell/4[$ and $-1$ if $x\in [(2n+1)\ell/4,(2n+1)3\ell/4[$, with integers $n$. Hence, the rotation operator (\ref{app:eq:Rotations}) reads:
\begin{align}
e^{i\frac{\phi}{2}\hat\Gamma^1_z}&=\int_{-\infty}^\infty  dx\  e^{i\frac{\phi}{2}s_z(x) }\ket{x}_x\bra{x}_x,
\label{app:eq:RotationzPos}
\end{align} 
which is a simple position phase gate that can be implemented through the SLM operation $\hat U_{\text{SLM}}$ with $h(x)=\frac{\phi}{2}s_z(x)$. 
Similarly, we can write 
\begin{align}
\hat\Gamma^1_ x&=\int_{-\ell/4}^{\ell/4}d\bar x \int_{-\pi/\ell}^{\pi/\ell} d\bar p\ \hat\sigma_x(\bar x,\bar p) \nonumber \\
&=\int_{-\ell/4}^{3\ell/4}d\bar x \int_{-\pi/\ell}^{\pi/\ell} d\bar p s_x(\bar x,\bar p) \ket{\bar x,\bar p}\bra{\bar x,\bar p},
\label{app:eq:Gamma1x}
\end{align}
where $s_x(\bar x,\bar p)$ takes the value $e^{i\bar p \ell/2}$ if $\bar x\in[-\ell/4,\ell/4[$ or $e^{i\bar p \ell/2}$ if $\bar x\in[\ell/4,3\ell/4[$. And again, by following the arguments in Appendix~\ref{app:Conditions}, we find that Eq.~(\ref{app:eq:Gamma1x}) can be written in the momentum representation as
\begin{align}
\hat\Gamma^1_x&=\int_{-\infty}^\infty dp \   s_x(p) \ket{p}_p\bra{p}_p,
\label{app:eq:GammaXMom}
\end{align} 
where $s_x(p)$ is a $4\pi/\ell$-periodic rectangular function taking the value $+1$ if $p\in [-(2m)\pi/\ell,(2m)\pi\ell/4[$ and $-1$ if $p\in [(2m+1)\pi/\ell,(2m+1)3\pi/\ell[$, with integers $m$. From Eq. (\ref{app:eq:GammaXMom}) then follows directly
\begin{align}
e^{i\frac{\phi}{2}\hat\Gamma^1_x}&=\int_{-\infty}^\infty dp \   e^{i\frac{\phi}{2}s_x(p) }\ket{p}_p\bra{p}_p,
\label{app:eq:RotationxPos}
\end{align} 
which is a position phase gate that can be implemented with a SLM operation programmed with the function $h(x)=\frac{\phi}{2}s_x(x)$, sandwiched between two Fourier transforms, as discussed previously. 

\subsection{Measuring the readout observables $\hat\Gamma_\beta$}\label{sec:MeasureGamma}

\subsubsection{Direct measurement}  
We first recall that the observables $\hat\Gamma_\beta$, with $\beta=x,y,z$,  correspond to phase space operators $F(\hat x,\hat p)$ fulfilling certain periodicity constraints, as discussed in the Appendix~\ref{app:Conditions}.  If we further consider only those readout observables that can be expressed as a function of a general quadrature $\hat x_\phi=\sin{(\phi)}\hat x+ \cos{(\phi)} \hat p$, we can write them in the corresponding diagonal form 
\begin{align}
F_\beta(\hat x_\phi)=\int_{-\infty}^\infty dx F_\beta(x) \ket{x}_{\phi}\bra{x}_{\phi}
\label{app:eq:Fdiagquad}
\end{align}
where the subscript denotes the $\hat x_\phi$-representation. Examples, as mentioned previously, are $\hat\Gamma_x=\cos{(\hat p \ell/2)}$, $\hat\Gamma_z=\cos{(2\pi \hat x/\ell)}$ and $\hat\Gamma_y=\cos{(2\pi \hat x/\ell-\hat p\ell/2)}=\cos{[\frac{2\pi}{\ell}g(\sin{(\phi')} \hat x+\cos{(\phi')}\hat p)]}$, being functions of $\hat x_{\frac{\pi}{2}}$, $\hat x_{0}$ and $\hat x_{\phi'}$, where $g=\sqrt{1+\ell^4/(4\pi)^2}$ and $\phi'=\arctan{[-\ell^2/(4\pi)]}$. 
 Accordingly, the expectation value of the operator (\ref{app:eq:Fdiagquad}) reads:
\begin{align}
\expec{F_\beta(\hat x_\phi)}=\int_{-\infty}^\infty dx F_\beta(x)\  | \bra{x}_{\phi}\ket{\Psi}|^2,
\label{app:eq:expecFquad}
\end{align}
which is solely determined by the probability density $p_{\phi}(x)=| \bra{x}_{\phi}\ket{\Psi}|^2$. We can reproduce the same reasoning in a bipartite system where we get for a product of two readout observables $\hat\Gamma_\beta \otimes \hat\Gamma_{\beta'}$:  
\begin{align}
F_\beta(\hat x_{\phi_1}) \otimes F_{\beta'}(\hat x_{\phi_2})=&\iint_{-\infty}^\infty dx_1dx_2 F_\beta(x_2)F_\beta(x_1) \nonumber \\ 
&\times \ket{x_1}_{\phi}\ket{x_2}_{\theta}\bra{x_1}_{\phi_1}\bra{x_2}_{\phi_2}
\label{app:eq:FdiagquadBipartite}
\end{align}
and the corresponding expectation value:
\begin{align}
\expec{F_\beta(\hat x_{\phi_1}) \otimes F_{\beta'}(\hat x_{\phi_2})}=&\iint_{-\infty}^\infty dx_1dx_2 F_\beta(x_2)F_\beta(x_1) \nonumber \\ 
&\times |\bra{x_1}_{\phi_1}\bra{x_2}_{\phi_2}\ket\Psi |^2,
\label{app:eq:expecFquadBipartite}
\end{align}
with the joint-probability density $p_{\phi_1,\phi_2}(x_1,x_2)=| \bra{x_1}_{\phi_1}\bra{x_2}_{\phi_2}\ket{\Psi}|^2$.

In an experimental setup with pairs of single photons we can determine the position or momentum probability densities $p_{0}(x)$ or $p_{\frac{\pi}{2}}(p)$, by detecting the position of the photons in the near- or far-field with respect to the output plane of the source of the photons. Position measurements of single photons can be performed either by scanning a single photon counter in the transverse plane of the photon or by using a single-photon sensitive camera \cite{Lantz,Camera1}. Arbitrary quadratures $\hat x_\phi$ can be assessed via fractional Fourier transforms realized with lens systems \cite{FracFour1,FracFour2}, allowing to determine arbitrary distributions $p_{\phi}(x)$. Finally, we can use Eq.~(\ref{app:eq:expecFquad}) to calculate expectation values of the desired readout observables $\hat\Gamma_\beta$.

The same measurement schemes can be applied to entangled pairs of photons (see Sec. \ref{sec:CreationWaveFkt}), using respectively two single photon counters or two single photon sensitive cameras, in order to determine the joint-probability distributions $p_{\phi_1,\phi_2}(x_1,x_2)$.

\subsubsection{Indirect measurement}  \label{sec:MeasurementPOVM}
\begin{figure}[t]
\includegraphics[scale=0.55]{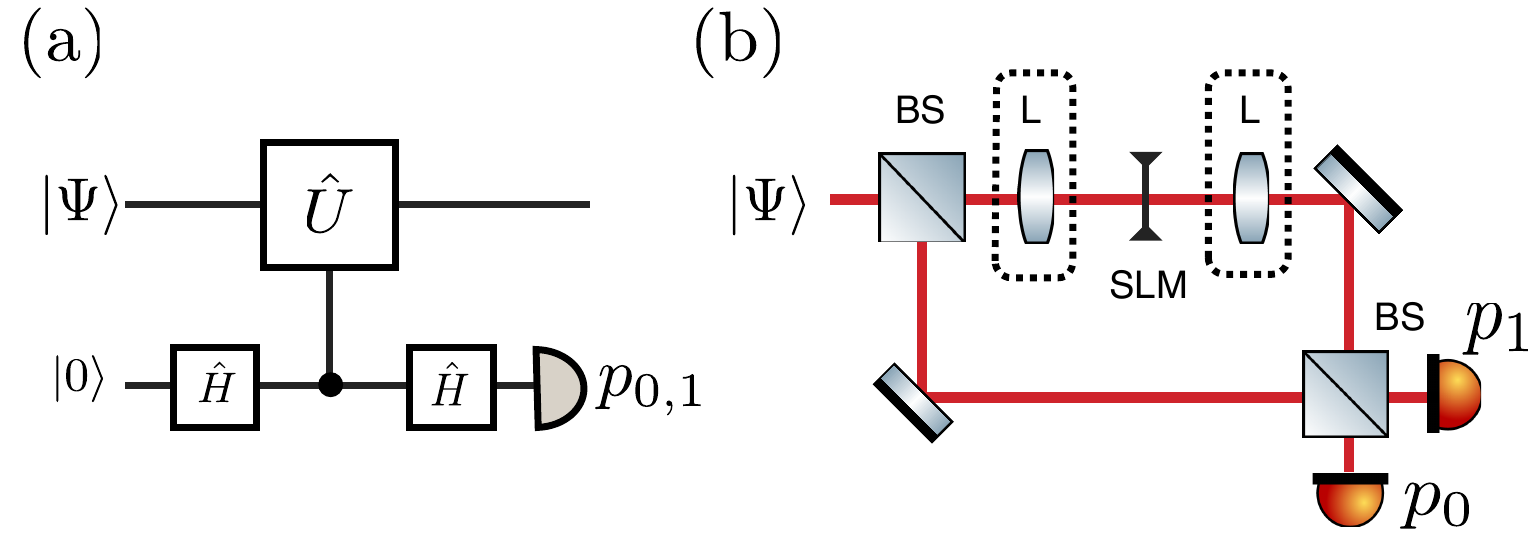}
\caption{(color online) (a) Quantum circuit allowing for the measurement of the observables (\ref{eq:Gammas}). $\hat H$ depict Hadamard gates and a controlled unitary gate $\hat U$ is applied if the ancilla is in the state $\ket 0$. The expectation value of (\ref{eq:Gammas}) is given by $p_0-p_1$, where $p_{0}$ ($p_{1}$) are the probabilities of detecting the ancilla in the state $\ket{0}$ ($\ket{1}$). In the case of the specific example mentioned in the text we choose $\hat U=\hat X$, $\hat Y$, $\hat Z$. (b) Proposal of an experimental implementation of circuit (a) using the spatial field of  single photons passing through a Mach-Zehnder interferometer. Controlled unitaries are realized by linear optical transformations inserted in one arm of the interferometer. Unitaries of the form $e^{i h(\hat x)}$ or $e^{i h(\hat p)}$ can be implemented using a SLM and lenses (\sffamily L\normalfont) allowing us to switch from the position to the momentum space.}
\label{fig_4}
\end{figure}

The expectation values of the observables $\hat\Gamma_\beta$, $\beta=x,y,z$,  can also be measured indirectly. Figure~\ref{fig_4} (a) shows the quantum circuit that allows for a indirect measurement of the expectation values of (\ref{eq:Gammas}) in their general form by coupling the CV system to an ancilla qubit and performing controlled unitary operations. In the following, we assume that the spectrum of the operators $\hat\Gamma_\beta$ is bounded by one. 
Let us define the following POVM elements ({effects}): 
\begin{align}
\hat E_+&=\frac{1}{2}(\mathbb 1+\hat\Gamma_\beta) \\
\hat E_-&=\frac{1}{2}(\mathbb 1- \hat\Gamma_\beta)
\end{align}
which satisfy the relation $\hat E_++\hat E_-=\mathbb 1$. The probability to obtain the outcome $+$ or $-$ is thus given by $p_{+}=\expec{ \hat E_{+}}$ or $p_{-}=\expec{ \hat E_{-}}=1-p_+$, respectively, and we can calculate $\expec{\hat\Gamma_\beta}=\expec{\hat E_+}-\expec{\hat E_-}=p_+-p_-$. Hence, the expectation value of every $\hat\Gamma_\beta$ can always be measured in terms of a two-valued POVM. More generally, if the spectrum of $\hat\Gamma_\beta$ is bounded between $\gamma_-$ and $\gamma_+$, one can simply rescale the spectrum of the corresponding POVM to reproduce the same argument \cite{Horodecki}.

Consider the quantum circuit shown in Fig.~\ref{fig_4}(a) which implements the operation
\begin{align}
\ket\Psi \ket 0 \rightarrow   \frac{1}{2}(\mathbb 1+\hat U) \ket\Psi\ket 0 +  \frac{1}{2}(\mathbb 1-\hat U)\ket\Psi\ket 1
\end{align}
on the initial state $\ket\Psi \ket 0$. Hence, by measuring the ancilla state in the basis $\ket{0/1}$, we project the system state onto $\ket{\Psi_\pm}=\hat D_\pm\ket\Psi=\frac{1}{2}(\mathbb 1\pm\hat U)\ket\Psi$ with the probability $p_{0/1}=\braket{\Psi_\pm}{\Psi_\pm}=\bra{\Psi}\hat D_\pm^\dagger \hat D_\pm\ket{\Psi}=\bra{\Psi}\hat E_\pm\ket{\Psi}$, which is equivalent to measuring the POVM $\hat E_\pm$ with the corresponding effects $\hat D_\pm$. With a general unitary operator $\hat U=e^{i g(\hat x,\hat p)}$, where $g(\hat x,\hat p)$ is a real function of position and momentum operator, we can also write $\hat E_\pm=\frac{1}{2}\{\mathbb 1\pm \cos{[g(\hat x,\hat p)}]\}$, leading to $p_+-p_-=\expec{\cos{[g(\hat x,\hat p)}]}$. Now, in order to measure any of the observables $\hat\Gamma_\beta$, we define $g(\hat x,\hat p)=\arccos{(F(\hat x,\hat p))}$, with the corresponding phase-space operator $F_\beta(\hat x,\hat p)$ (see Appendix~\ref{app:Conditions}), yielding $p_+-p_-=\expec{F_\beta(\hat x,\hat p)}=\expec{\hat\Gamma_\beta}$.

The above measurement strategy can be straightforwardly implemented with single photons passing through balanced Mach-Zehnder interferometers, as depicted in Fig.~\ref{fig_4}(b). Therein, the spatial distribution of the single photons represent the CV system and the path of the interferometer the state of the ancilla. Controlled unitary operations are realized via linear optical elements placed in one of the arms of the interferometer, and measurements of the ancilla state by detecting photons that exit form one of the two output ports using single photon bucket detectors. 

A SLM, with the option of additionally placing it in the Fourier plane between to lenses, allows us to perform arbitrary position or momentum phase gates, $e^{ih(\hat x)}$ or $e^{ih(\hat p)}$, where $h(\cdot)$ is user defined on the SLM. As discussed previously $x$ and $p$ can be considered as the near- and far-field variables with respect to the output plane of the source.  Phase gates $e^{ih(\hat x_\phi)}$ in terms of an arbitrary quadrature $\hat x_\phi=\sin{(\phi)} \hat x+\cos{(\phi)} \hat p$ can be realized through fractional Fourier transform before and after the SLM \cite{FracFour1,FracFour2}. Hence, we have the ability to implement a broad class of unitaries on the spatial distribution of the photons allowing us to measure expectation values through $p_+-p_-=\expec{\cos{[h(\hat x_\phi)]}}$, yielding $\expec{\hat\Gamma_\beta}$.

At this point we note that the direct measurement of the observables $\hat\Gamma_\beta$, as described in this section, is less expensive in terms of the number of measurements needed  in order to determine the expectation values $\expec{\hat\Gamma_\beta}$, as compared to the indirect measurement strategy introduced in the previous section \cite{interferometer1,interferometer2}. Moreover, there is no need for a spatially resolving detection of the photons.

\section{Conclusion and perspectives} \label{sec:Conclusion}
We presented a general framework which allows us to encode, manipulate and readout discrete quantum information in phase space in terms of continuous variables states. This is possible by using the modular variables formalism that naturally leads to an intuitive definition of a qubit and the necessary universal manipulations. We demonstrate its strong relationship with the GKP formalism, and show that as far as one is interested in performing quantum protocols involving expectation values of bounded periodic observables, so called modular variables, it is possible to encode binary quantum information in more general states than the GKP ones. A possible experimental implementation using the transverse degrees of freedom of single photons was discussed, as well.

From a fundamental point of view, our framework shows how to reveal naturally discrete structures of states and operations written in a continuous variable representation. Furthermore, it provides a unifying formalism that shows how, in general, measurements of modular variables can be employed in quantum information protocols. Finally, an application of our ideas in hybrid quantum systems, which use CV besides some discrete degree of freedom, as is the case for single photons, could be advantageous for future experimental implementations of quantum information protocols.
\begin{acknowledgments}
The authors are indebted to F. de Melo, O. Jim\'enez Far\'ias, L. Aolita, A. D. Ribeiro, N. Menicucci, R. N. Alexander, G. Ferrini, A. Laversanne-Finot and T. Douce for inspiring discussions. The authors acknowledge financial support by ANR/CNPq HIDE,  ANR COMB, FAPERJ, CNPq and CAPES/COFECUB project Ph 855-15.
\end{acknowledgments}
\appendix
\section{Identifying qubits in the modular variables representation}\label{app:LogicalQubits}
To show, that every state in the modular representation can indeed be written as in Eq. (\ref{eq:GeneralState}), we start from Eq. (\ref{app:eq:ModVarState}) and split the integration over $\bar x$ into two equally sized domains:
\begin{align}
\ket{\Psi}=\int_{-\ell/4}^{\ell/4}d\bar x &\int_{-\pi/\ell}^{\pi/\ell} d\bar p \ \psi(\bar x,\bar p) \ket{\bar x,\bar p} \nonumber \\
&+ \psi(\bar x+\ell/2,\bar p)\ket{\bar x+\ell/2,\bar p}.
\label{eqn:arbState}
\end{align}
Next, we redefine the complex {modular wave function} $\psi(\bar x,\bar p)=|\psi(\bar x,\bar p)|e^{i \varphi(\bar x,\bar p)}$ in the following way:
\begin{align}
\psi(\bar x,\bar p)&:= f(\bar x,\bar p) \cos{\left(\theta(\bar x,\bar p)/2\right)},  \\
\psi(\bar x+\ell/2,\bar p)&:=f(\bar x,\bar p) e^{i\phi(\bar x,\bar p)} \sin{\left(\theta(\bar x,\bar p)/2\right)}.
\label{eqn:representg}
\end{align}
where $f(\bar x,\bar p)=|f(\bar x,\bar p)|e^{i\epsilon(\bar x,\bar p)}$ is complex amplitude, and $\epsilon(\bar x,\bar p)$, $\theta(\bar x,\bar p) $ and $\phi(\bar x,\bar p)$ are real functions, all defined on the domain $[0,\ell/2[$. These functions directly relate to the {modular wave function} $\psi(\bar x,\bar p)$ through:
\begin{align}
|f(\bar x,\bar p)|&=\sqrt{|\psi(\bar x,\bar p)|^2+|\psi(\bar x+\ell/2,\bar p)|^2}, \\
\epsilon(\bar x,\bar p)&=\varphi(\bar x,\bar p),
\label{eqn:gfcorresp1}
\end{align}
and
\begin{align}
\theta(\bar x,\bar p)&=2\ \text{arccot}{\left|\frac{\psi(\bar x,\bar p)}{\psi(\bar x+\ell/2,\bar p)}\right| }, \\
\phi(\bar x,\bar p)&=\varphi(\bar x+\ell/2,\bar p)-\varphi(\bar x,\bar p).
\label{eqn:gfcorresp2}
\end{align}
Hence, we can express every state as:
\begin{align}
\ket{\Psi}=\int_{-\ell/4}^{\ell/4}d\bar x &\int_{-\pi/\ell}^{\pi/\ell} d\bar p f(\bar x,\bar p) \ket{{\Psi}(\bar x,\bar p)},
\label{app:eq:ContQubitRep1}
\end{align}
with
\begin{align}
\ket{{\Psi}(\bar x,\bar p)}&=\cos{(\frac{\theta(\bar x,\bar p)}{2})} \ket{\bar x,\bar p} \nonumber \\
&+ \sin{(\frac{\theta(\bar x,\bar p)}{2})} e^{i\phi(\bar x,\bar p)} \ket{\bar x+\ell/2,\bar p}.
\label{app:eq:ContQubitRep2}
\end{align}
Later on we will, for technical reasons, express certain operators in a slightly modified basis defined as $\{\ket{|{\bar x,\bar p}}, \ket{|{\bar x+\ell/2 ,\bar p}}\}=\{e^{-i\bar p \ell/4}\ket{\bar x,\bar p}, e^{i\bar p \ell/4}\ket{\bar x+\ell/2 ,\bar p}\}$, which involve additional $\bar p$-dependent phase factors which can be absorbed in the definition of the wave function $f(\bar x,\bar p)$. In the case of the above discussed example of a comb of Gaussian spikes with Gaussian enelope $f(\bar x,\bar p)$ becomes a Gaussians with periodic boundary conditions. 

Note that it is equivalently possible to define such a qubit structure in terms of the modular momentum $\bar p$. In this case, one splits the integration over $\bar p$, in Eq. (\ref{eqn:arbState}), into two parts and obtains a similar result to Eqs. (\ref{app:eq:ContQubitRep1}) and (\ref{app:eq:ContQubitRep2}), now with
\begin{align}
\ket{{\Psi_p}(\bar x,&\bar p)}=\cos{(\frac{\theta_p(\bar x,\bar p)}{2})} \ket{\bar x,\bar p} \nonumber \\
&+ \sin{(\frac{\theta_p(\bar x,\bar p)}{2})} e^{i\phi_p(\bar x,\bar p)} \ket{\bar x,\bar p+\pi/\ell}.
\label{app:eq:ContQubitRep3}
\end{align}
where the $p$-subscripts in Eq. (\ref{app:eq:ContQubitRep3}) refer to the splitting with respect to $\bar p$. Furthermore, this intrinsic qubit structure can be generalized to qudit systems by splitting the integration in Eq. (\ref{eqn:arbState}) into $d$-parts instead of two, as discussed also in \cite{Pierre}. However, in the context of this work we will restrict ourselves to the above presented case of $d=2$.

\section{State readout with modular variables} 
\subsection{Definition and important relations}\label{app:GammaProp}
The readout of the encoded logical state can be performed using the observables defined in Eq. (\ref{eq:Gammas}) of the main text. For convenience, we reproduce their definition here:
\begin{align}
\hat{\Gamma}_\beta=\int_{-\ell/4}^{\ell/4}d\bar x \int_{-\pi/\ell}^{\pi/\ell} d\bar p\  \zeta_\beta (\bar x, \bar p) \hat\sigma_\beta(\bar x,\bar p),
\label{app:eq:Gammas}
\end{align}
where 
\begin{align}
\hat\sigma_x(\bar x,\bar p)&= \ket{|{\bar x,\bar p} }  \bra{{\bar x+\ell/2,\bar p}|}+ \ket{|{\bar x,\bar p}}  \bra{{\bar x+\ell/2,\bar p}|}, \label{app:eq:sigx}\\
\hat\sigma_z(\bar x,\bar p)&= \ket{|{\bar x,\bar p} }  \bra{{\bar x,\bar p}|} - \ket{|{\bar x+\ell/2,\bar p} }  \bra{{\bar x+\ell/2,\bar p}|},\label{app:eq:sigz} \\
\hat\sigma_y(\bar x,\bar p)&=i( \ket{|{\bar x,\bar p} }  \bra{{\bar x+\ell/2,\bar p}|}- \ket{|{\bar x,\bar p}}  \bra{{\bar x+\ell/2,\bar p}|}),\label{app:eq:sigy}
\end{align}
where $\zeta_\beta(\bar x,\bar p)$ are real functions defined on the domain $[-\ell/4,\ell/4[\times[-\pi/\ell,\pi/\ell[$. In Sec.~\ref{app:Conditions} we will show which class of general phase space operators $F(\hat x,\hat p)$ obey such a representation, however, for the moment we take their form as granted to discuss several important properties. First, we note that the matrix elements of the operators (\ref{app:eq:sigx}), (\ref{app:eq:sigz}) and (\ref{app:eq:sigy}) read:
\begin{widetext}
\begin{align}
\langle \bar x', \bar p'| \hat\sigma_{z}(\bar x_0, \bar p_0) |\bar x, \bar p\rangle&=\delta(\bar p-\bar p_0) \delta(\bar p'-\bar p_0) \nonumber \\
&\times \left[\delta(\bar x-\bar x_0)\delta(\bar x'-\bar x_0)
-\delta(\bar x-\frac{\ell}{2}-\bar x_0)\delta(\bar x'-\frac{\ell}{2}-\bar x_0)\right], \label{eqapp:SigZxpMatEl}\\
\langle\bar x', \bar p' | \hat\sigma_{x}(\bar x_0, \bar p_0) |\bar x, \bar p\rangle&=\delta(\bar p-\bar p_0) \delta(\bar p'-\bar p_0) \nonumber \\
&\times \left[\delta(\bar x'-\bar x_0)\delta(\bar x-\frac{\ell}{2}-\bar x_0)e^{-i\bar p\ell/2}
+\delta(\bar x'-\frac{\ell}{2}-\bar x_0)\delta(\bar x-\bar x_0)e^{i\bar p\ell/2}\right],\label{eqapp:SigXxpMatEl} \\
\langle\bar x', \bar p' | \hat\sigma_{y}(\bar x_0, \bar p_0) |\bar x, \bar p\rangle&=-i\delta(\bar p-\bar p_0) \delta(\bar p'-\bar p_0) \nonumber \\
&\times\left[\delta(\bar x'-\bar x_0)\delta(\bar x-\frac{\ell}{2}-\bar x_0)e^{-i\bar p\ell/2}
-\delta(\bar x'-\frac{\ell}{2}-\bar x_0)\delta(\bar x-\bar x_0)e^{i\bar p\ell/2}\right].\label{eqapp:SigYxpMatEl}
\end{align}
\end{widetext}
Now, using Eqs.~(\ref{app:eq:sigx})-(\ref{eqapp:SigYxpMatEl}), we can show that the $(\bar x,\bar p)$-dependent Pauli matrices $\sigma_\alpha(\bar x,\bar p)$, with $\alpha=x,y,z$, fulfill the relation:
\begin{align}
&\hat\sigma_\beta (\bar x,\bar p)\hat\sigma_\beta (\bar x',\bar p')=\delta(\bar x'-\bar x)\delta(\bar p'-\bar p)  \nonumber \\
&\times \left[i\sum_{\gamma=x,y,z} \varepsilon_{\alpha\beta\gamma}   \hat\sigma_\gamma (\bar x,\bar p)+\delta_{\alpha,\beta} \mathbb 1 (\bar x,\bar p) \right]
\label{eq:xpPauliRelation}
\end{align}
where $\alpha,\beta=x,y,z$ and $\mathbb 1 (\bar x,\bar p)= \ket{|{\bar x,\bar p} }  \bra{{\bar x,\bar p}|} +\ket{|{\bar x+\ell/2,\bar p} }  \bra{{\bar x+\ell/2,\bar p}|}$. 
The relation (\ref{eq:xpPauliRelation}) resembles the one of a real Pauli algebra with additional $\delta$ functions ensuring that the products of Pauli operators corresponding to different subspaces, labeled by $(\bar x,\bar p)$ and $(\bar x',\bar p')$, respectively, vanish. Further on, we can calculate the expectation value of the observables (\ref{app:eq:Gammas}) with respect to an arbitrary CV state expressed in the modular representation (\ref{app:eq:approxGKP}), yielding:
\begin{widetext}
\begin{align}
\expec{{\hat\Gamma}_{x}} =
&\iiint_{-\ell/4}^{\ell/4}d\bar x\ d\bar x_1d\bar x_2 \iiint_{-\pi/\ell}^{\pi/\ell} d\bar p\ d\bar p_1d\bar p_2 \ \zeta_x (\bar x, \bar p) f^*(\bar x_1, \bar p_1) f(\bar x_2, \bar p_2 ) \nonumber \\
&\times  \bra{\Psi(\bar x_1, \bar p_1)}\hat\sigma_{x}(\bar x, \bar p)\ket{\Psi(\bar x_2, \bar p_2)}   \nonumber \\
=&\int_{-\ell/4}^{\ell/4}d\bar x \int_{-\pi/\ell}^{\pi/\ell} d\bar p \   \zeta_x (\bar x, \bar p) |f(\bar x, \bar p )|^2 \sin{\theta (\bar x,\bar p)}  \cos{\phi (\bar x,\bar p)},
\label{eqapp:expecGammaX}
\end{align}
\end{widetext}
In the second step of the computation (\ref{eqapp:expecGammaX}) we dropped terms that are proportional to cross products of $\delta$ functions as, for instance, $\delta(\bar x_1+\ell/2-\bar x)\delta(\bar x-\bar x_2)$, because, upon integration of $\bar x_1$ and $\bar x_2$ over the interval $[-\ell/4,\ell/4[$, such terms are only nonzero in a single point being a set of measure zero, and thus the integration vanishes. Equivalently, the expectation values of the observables $\hat\Gamma_y$ and $\hat\Gamma_z$ read: 
\begin{align}
\expec{{\hat\Gamma}_{y}} =&\int_{-\ell/4}^{\ell/4}d\bar x \int_{-\pi/\ell}^{\pi/\ell} d\bar p \ \zeta_y (\bar x, \bar p) |f(\bar x, \bar p )|^2 \nonumber \\
&\times \sin{\theta (\bar x,\bar p)}  \sin{\phi (\bar x,\bar p)}, \label{eq:expecGammaY}\\
\expec{{\hat\Gamma}_{z}}=&\int_{-\ell/4}^{\ell/4}d\bar x \int_{-\pi/\ell}^{\pi/\ell} d\bar p \  \beta_z (\bar x, \bar p) |f(\bar x, \bar p )|^2 \nonumber \\
&\times \cos{\theta (\bar x,\bar p)}.
\label{eq:expecGammaZ}
\end{align}
Or in vector notation we can write 
\begin{align}
\expec{\hat{\boldsymbol\Gamma}} =\int_{-\ell/4}^{\ell/4}d\bar x \int_{-\pi/\ell}^{\pi/\ell} d\bar p \   \zeta_\beta (\bar x, \bar p) |f(\bar x, \bar p )|^2 \boldsymbol\vartheta (\bar x,\bar p)
\end{align}
with 
\begin{align}
\boldsymbol\vartheta (\bar x,\bar p)=&(\vartheta_x (\bar x,\bar p),\vartheta_y (\bar x,\bar p),\vartheta_z (\bar x,\bar p)) \nonumber \\
=&(\sin{\theta(\bar x,\bar p)}\cos{\phi(\bar x,\bar p)},\nonumber \\
&\sin{\theta(\bar x,\bar p)}\sin{\phi(\bar x,\bar p)} ,\cos{\theta(\bar x,\bar p)}).
\end{align}
Further on, by summing over the squares of the expectation values (\ref{eqapp:expecGammaX}), (\ref{eq:expecGammaY}) and (\ref{eq:expecGammaZ}) we can show:
\begin{widetext}
\begin{align}
\expec{\hat{\boldsymbol\Gamma}}^2=&\expec{{\hat\Gamma}_{x}}^2+\expec{{\hat\Gamma}_{y}}^2+\expec{{\hat\Gamma}_{z}}^2 \nonumber \\
=&\iint_{-\ell/4}^{\ell/4}d\bar x d\bar x' \iint_{-\pi/\ell}^{\pi/\ell} d\bar p d\bar p' \sum_{\beta=x,y,z}    |f(\bar x, \bar p )|^2 |f(\bar x', \bar p' )|^2   \zeta_\beta (\bar x, \bar p)  \zeta_\beta (\bar x', \bar p') \nonumber \\
&\times v_\beta (\bar x,\bar p) v_\beta (\bar x',\bar p') \nonumber  \\
\leq &\iint_{-\ell/4}^{\ell/4}d\bar x \bar x' \iint_{-\pi/\ell}^{\pi/\ell} d\bar p d\bar p' \sum_{\beta=x,y,z}    |f(\bar x, \bar p )|^2 |f(\bar x', \bar p' )|^2   \zeta_\beta (\bar x, \bar p)  \zeta_\beta (\bar x', \bar p') \nonumber \\
&\times \frac{1}{2}[v_\beta (\bar x,\bar p)^2+ v_\beta (\bar x',\bar p')^2] \nonumber  \\
\leq &\big( \max_{\bar x,\bar p,\beta}|\zeta_\beta{(\bar x,\bar p)}|\big)^2  \iint_{-\ell/4}^{\ell/4}d\bar x \bar x' \iint_{-\pi/\ell}^{\pi/\ell} d\bar p d\bar p'  |f(\bar x, \bar p )|^2 |f(\bar x', \bar p' )|^2  \nonumber \\
&\times\underbrace{ \frac{1}{2}\left[\sum_{\beta=x,y,z}v_\beta^2 (\bar x,\bar p)+\sum_{\beta=x,y,z} v_\beta^2 (\bar x',\bar p')  \right]}_{=1} \nonumber \\
=& \big( \max_{\bar x,\bar p,\beta}|\zeta_\beta{(\bar x,\bar p)}|\big)^2 \left(\int_{-\ell/4}^{\ell/4}d\bar x \int_{-\pi/\ell}^{\pi/\ell} d\bar p \ |f (\bar x, \bar p)|^2 \right)^2  
\leq \big( \max_{\bar x,\bar p,\beta}|\zeta_\beta{(\bar x,\bar p)}|\big)^2,
\label{eqn:expecGammaApp}
\end{align}
\end{widetext}
where we use that $[\vartheta_\beta (\bar x,\bar p)-\vartheta_\beta (\bar x',\bar p')]^2\geq 0$ and that the Bloch vector of a pure qubit state is normalized to $1$. For the example discussed in the main text, we  have $\max_{\bar x,\bar p, \beta}|\zeta_\beta{(\bar x,\bar p)}|= 1$, which shows that $\expec{\hat{\boldsymbol\Gamma}}$ is contained in a unit sphere. 

\subsection{Conditions on phase space operators to obey the form (\ref{app:eq:Gammas})}\label{app:Conditions}
Let's consider an arbitrary observable in phase space, \textit{i.e.} a valid function of the position and momentum operator, expressed in the modular basis:
\begin{align} 
F(\hat x,\hat p)&= 
\iint_{0}^{\ell}  d\bar x d\bar x'  \iint_{0}^{\frac{2\pi}{\ell}} d\bar p d\bar p'  \nonumber \\
&\times \underbrace{\langle\bar x, \bar p|  F(\hat x,\hat p) |\bar x', \bar p'\rangle}_{\equiv f(\bar x, \bar p;\bar x', \bar p')} |\bar x, \bar p\rangle \langle\bar x', \bar p'| .
\label{app:eq:GenModVarOp1}
\end{align}
with the matrix elements $f(\bar x, \bar p;\bar x', \bar p')$, which, by using the definition of the modular eigenstates in Eq.~(\ref{app:eq:ModVarEigenstates}), can be expressed as
\begin{align}
F(\bar x, \bar p;\bar x', \bar p')=& \frac{\ell}{2\pi} \sum_{\nu,\mu=-\infty}^{\infty} e^{\frac{i}{\hbar}(\bar{p}'\mu-\bar{p}\,\nu)\ell} \nonumber \\
&\times \langle\bar x+ \nu \ell |F(\hat x,\hat p)|\bar{x}'+ \mu \ell \rangle.
\label{app:eq:GenModVarOp2}
\end{align}
Further on, if we assume that the function $F(\hat x,\hat p)$ is periodic with respect to $\hat x$ and $\hat p$ with period $\ell$ and $2\pi/\ell$, respectively, we can rewrite it as a double Fourier series:
\begin{align} 
F(\hat x,\hat p)&= \sum_{n=-\infty}^\infty  \sum_{m=-\infty}^\infty  d_{n,m} e^{2\pi i n \hat x/L-iL' m \hat p}
\label{app:eq:DoubleFourier}
\end{align}
where $d_{n,m}$ are the complex Fourier coefficients obeying the normalization condition $ \sum_{n=-\infty}^\infty  \sum_{m=-\infty}^\infty  |d_{n,m}|^2=1$, and we have, by definition, $F( x+L', p+2\pi/L)=F( x, p)$. In the following, we discuss how one can construct the observables (\ref{app:eq:Gammas}) from periodic operators of the form $F(\hat x,\hat p)$.

\subsubsection{$\hat\Gamma_z$-operator} 
To start, we assume a modular operator that is diagonal in the modular position and momentum with the following matrix elements:
\begin{eqnarray}
F_z(\bar x, \bar p;\bar x', \bar p')= \delta(\bar x-\bar x')\,\delta(\bar p-\bar p') \ 
\tilde F_z(\bar x, \bar p), 
\label{eq:FzDiagEl}
\end{eqnarray}
which fulfill the periodicity condition $\tilde F_z(\bar x, \bar p)=-\tilde F_z(\bar x-\frac{l}{2}, \bar p)$. Then we obtain
\begin{align}
F_z&(\hat x,\hat p) =\int_{-\ell/4}^{\ell/4} d\bar x \int_{-\pi/\ell}^{\pi/\ell} d\bar p  \, \tilde F_z(\bar x, \bar p) \nonumber \\
&\times (  |\bar x, \bar p\rangle\langle\bar x, \bar p| -
  |\bar x+\frac{\ell}{2}, \bar p\rangle\langle\bar x+\frac{\ell}{2}, \bar p| ) \nonumber \\
  &=\int_{-\ell/4}^{\ell/4} d\bar x \int_{-\pi/\ell}^{\pi/\ell} d\bar p  \, \tilde F_z(\bar x,\bar p) \hat\sigma_z(\bar x, \bar p) \equiv \hat\Gamma_z.
\end{align}
where we defined (for $-\frac{\ell}{4}\le\bar x <\frac{\ell}{4}$ and $-\frac{\pi}{\ell}\le\bar p <\frac{\pi}{\ell}$)
\begin{align}
\hat\sigma_z(\bar x, \bar p)&= e^{+i\theta_-}|\bar x, \bar p\rangle\langle\bar x, \bar p|e^{-i\theta_-}  \nonumber \\
&-e^{+i\theta_+}|\bar x+\frac{\ell}{2}, \bar p\rangle\langle\bar x+\frac{\ell}{2}, \bar p |e^{-i\theta_+} ,
\end{align}
with the phases $\theta_\pm=\theta_\pm(\bar x, \bar p)$ which, up to now, can assume any value. Now, if we assume a phase space operator of the form (\ref{app:eq:DoubleFourier}), with $L=L'=\ell$, and use Eq. (\ref{app:eq:GenModVarOp2}) we get
\begin{align}
F_z(\bar x, \bar p;\bar x', \bar p')=&\frac{\ell}{2\pi}\sum_{\nu,\mu,n,m=-\infty}^{\infty}e^{i(\bar{p}'\mu-\bar{p}\nu)\ell} d_{n,m} \nonumber \\
&\times \bra{\bar x+\nu \ell}e^{2\pi i n \hat x/\ell-i \ell m \hat p}\ket{\bar x'+\mu \ell} \nonumber \\
=&\tilde F(\bar x,\bar p)  \delta(\bar x-\bar x') \delta(\bar p-\bar p').
\end{align}
where we used $\sum_{\mu=-\infty}^{\infty}e^{i(\bar{p}'-\bar{p})\mu \ell}=\frac{2\pi}{\ell} \delta(\bar p-\bar p')$. We thus find that periodic phase space operators with periodicity $\ell$  and $2\pi/\ell$ in $\hat x$ and $\hat p$, respectively, lead to diagonal operators in the modular representation with matrix elements $\tilde F(\bar x,\bar p)=\sum_{n,m=-\infty}^{\infty} d_{n,m}e^{2\pi i n \bar x-i\bar p m \ell}  e^{-i\pi n m}$. Moreover, to obtain the operator $\hat\Gamma_z$, the condition $\tilde F(\bar x+\ell/2,\bar p)=-\tilde F(\bar x,\bar p)$ needs to be enforced as well. The latter is true if $d_{n,m}=0$, for all even $n$, leading to the following form of the diagonal elements in Eq.~(\ref{eq:FzDiagEl}):
\begin{eqnarray}
\tilde F_z(\bar x,\bar p)=\sum_{n',m=-\infty}^{\infty} d_{2n'+1,m}e^{2\pi i (2n'+1) \bar x'/\ell-i\bar p m \ell}  e^{-i\pi m},\nonumber \\
\label{eq:F_z}
\end{eqnarray}
which correspond to the phase space observable:
\begin{align} 
F_z(\hat x,\hat p)&=\sum_{n',m=-\infty}^{\infty} d_{2n'+1,m} \hat D(\ell m,2\pi (2n+1)/\ell).
\label{eq:FzObservable}
\end{align}
Hence, with Eq.~(\ref{eq:FzObservable}) we provide a specific class of modular variables which in the modular representation can be expressed in the form of $\hat\Gamma_z$, with $\zeta_z(\bar x,\bar p)$ chosen according to Eq.~(\ref{eq:F_z}). A particular example of the observable (\ref{eq:FzObservable}) is given by choosing only two nonzero coefficients $d_{+1,0}=1/2$ and $d_{-1,0}=1/2$, leading to $F_z(\hat x,\hat p)=\cos{(2\pi \hat x/\ell)}=\text{Re}(\hat Z)$ and $F_z(\bar x,\bar p)=\cos{(2\pi \bar x/\ell)}$.

\subsubsection{$\hat\Gamma_x$-operator}
Next, we consider a modular operator defined by the matrix elements
\begin{align}
&F_x(\bar x, \bar p;\bar x', \bar p')=\tilde F_x(\bar x, \bar p) \delta(\bar p-\bar p') \nonumber \\
&\times\left\{\begin{array}{llrl}
e^{+i\kappa_x(\bar x,\bar p)}\, \delta(\bar x-(\bar x'+\ell/2)), & \text{for}& -\ell/4&\le \bar{x} \le \ell/4 \\
e^{-i\kappa_x(\bar x,\bar p)}\, \delta(\bar x-(\bar x'-\ell/2)), & \mathrm{for}& \ell/4&\le \bar{x} \le 3\ell/4
\end{array}\right. ,
\end{align}
with the periodicity properties $\tilde F_x(\bar x+\frac{\ell}{2}, \bar p)=\tilde F_x(\bar x, \bar p)$ and $\kappa_x(\bar x+\frac{\ell}{2}, \bar p)=\kappa_x(\bar x, \bar p)$, leading to
\begin{align}
F_x&(\hat x,\hat p) =\int_{-\ell/4}^{\ell/4} d\bar x \int_{-\pi/\ell}^{\pi/\ell} d\bar p  \, \tilde F_x(\bar x, \bar p) \nonumber \\
 &\times ( e^{+i\kappa_x(\bar x,\bar p)} |\bar x, \bar p\rangle\langle\bar x+\frac{\ell}{2}, \bar p| +
e^{-i\kappa_x(\bar x,\bar p)}  |\bar x+\frac{\ell}{2}, \bar p\rangle\langle\bar x, \bar p| ) \nonumber \\
  &=\int_{-\ell/4}^{\ell/4} d\bar x \int_{-\pi/\ell}^{\pi/\ell} d\bar p  \, \tilde F_x(\bar x,\bar p) \hat\sigma_x(\bar x, \bar p) \equiv \hat\Gamma_x,
  \label{app:eq:GammaXcond}
\end{align}
 where we defined
 \begin{align}
\hat\sigma_x(\bar x, \bar p) =& e^{+i\theta_-}|\bar x, \bar p\rangle\langle\bar x+\frac{\ell}{2}, \bar p|e^{-i\theta_+} \nonumber \\
&+ e^{+i\theta_+}|\bar x+\frac{\ell}{2}, \bar p\rangle\langle\bar x, \bar p|e^{-i\theta_-} ,
\end{align}
with $\theta_+(\bar x,\bar p)-\theta_-(\bar x,\bar p) =\kappa_x(\bar x,\bar p)$. 

Further on, we want to find the conditions on the general periodic phase space operator (\ref{app:eq:DoubleFourier}), such that it can be brought in the form (\ref{app:eq:GammaXcond}).  Therefore, we assume $L=L'=\ell/2$ and $d_{n,m}=0$, for all even $m$, which yields:
\begin{align}
F_x(\bar x,& \bar p;\bar x', \bar p')=\frac{\ell}{2\pi}\sum_{\nu,\mu,n,m=-\infty}^{\infty}d_{n,2m+1}e^{i(\bar{p}'\mu-\bar{p}\nu)\ell}  \nonumber \\
&\times \bra{\bar x+\nu \ell}e^{4\pi i n \hat x/\ell-i \ell (2m+1) \hat p/2}\ket{\bar x'+\mu \ell} \nonumber \\
=&\frac{\ell}{2\pi}\sum_{\nu,\mu,n,m=-\infty}^{\infty}d_{n,2m+1}e^{i(\bar{p}'\mu-\bar{p}\nu)\ell} e^{i \pi n}e^{4\pi i n \bar x'/\ell}  \nonumber \\
&\times \Big[\braket{\bar x+\nu \ell}{\bar x'+\frac{\ell}{2}+(\mu+m) \ell} \Theta_1(\bar x')\ + \nonumber \\
&\braket{\bar x+\nu \ell}{\bar x'-\frac{\ell}{2}+(\mu+m+1) \ell} \Theta_2(\bar x') \Big],
\label{app:eq:MatrixEleX}
\end{align}
where we split up the domain of $\bar x'$ with the two rectangular functions $\Theta_1(\bar x)=\Theta(\bar x+\ell/4)-\Theta(\bar x-\ell/4)$ and  $\Theta_2(\bar x)=\Theta(\bar x-\ell/4)-\Theta(\bar x-3\ell/4)$, defined in terms of the Heaviside stepfunction $\Theta(\bar x')$. With this Eq. (\ref{app:eq:MatrixEleX}) becomes:
\begin{align}
F_x(\bar x,& \bar p;\bar x', \bar p')=\frac{\ell}{2\pi}\sum_{\nu,\mu,n,m=-\infty}^{\infty}d_{n,2m+1}e^{i(\bar{p}'\mu-\bar{p}\nu)\ell} e^{i \pi n}   \nonumber \\
&\times e^{4\pi i n \bar x'/\ell} \Big[\delta(\bar x-(\bar x'+\frac{\ell}{2})) \delta_{\nu,\mu+m} \Theta_1(\bar x')\  \nonumber \\
&+\delta(\bar x-(\bar x'-\frac{\ell}{2})) \delta_{\nu,\mu+m+1} \Theta_2(\bar x')  \Big] \nonumber \\
=&\tilde F_x(\bar x,\bar p) \delta(\bar p-\bar p') \nonumber \\
&\times\left\{\begin{array}{llrl}
e^{+i\bar p\ell/2}\, \delta(\bar x+\ell/2-\bar x'), & \mathrm{for}& -\frac{\ell}{4}&\le \bar{x} \le \frac{\ell}{4} \\
e^{-i\bar p\ell/2}\, \delta(\bar x-\ell/2-\bar x'), & \mathrm{for}& \frac{\ell}{4}&\le \bar{x} \le \frac{3\ell}{4}
\end{array}\right.  ,
\label{app:eq:MatrixEleX1}
\end{align}
where 
\begin{align}
\tilde F_x(\bar x,\bar p)=\sum_{n,m=-\infty}^{\infty} d_{n,2m+1} e^{i\pi n}e^{4\pi i n \bar x-i\bar p \frac{\ell}{2} (2m+1)}.
\label{eq:F_x}
\end{align}
We thus find that all operators of the form 
\begin{align} 
F_x(\hat x,\hat p)&=\sum_{n,m=-\infty}^{\infty} d_{n,2m+1} \hat D(\ell (2m+1)/2,4\pi n/\ell),
\label{eq:FxObservable}
\end{align}
where we set $L=L'=\ell/2$ and $d_{n,m}=0$, for all even $m$, can be expressed as $\hat\Gamma_x$ with $\zeta_x(\bar x,\bar p)=\tilde F_x(\bar x,\bar p)$ and $\kappa_x(\bar x,\bar p)=\bar p\ell/2$. An example of Eq.~(\ref{eq:FxObservable}) is given by the operator $\text{Re}(\hat X)=\cos{\hat p\ell/2}$, where only $d_{0,1}=1/2$ and $d_{0,-1}=1/2$ are nonzero, which is equal to $\hat\Gamma_x$ with $\zeta_x(\bar x,\bar p)=F_x(\bar x,\bar p)=\cos{\bar p\ell/2}$.

\subsubsection{$\hat\Gamma_y$-operator}
Finally, we consider a modular operator defined by the matrix elements
\begin{align}
&F_y(\bar x, \bar p;\bar x', \bar p')=\tilde F_y(\bar x, \bar p) \delta(\bar p-\bar p')  \nonumber \\
&\times\left\{\begin{array}{llrl}
(+i) e^{+i\kappa_y(\bar x,\bar p)}\, \delta(\bar x-(\bar x'+\ell/2)), & \mathrm{for}& -\frac{\ell}{4}&\le \bar{x} \le \frac{\ell}{4} \\
(-i) e^{-i\kappa_y(\bar x,\bar p)}\, \delta(\bar x-(\bar x'-\ell/2)), & \mathrm{for}& \frac{\ell}{4}&\le \bar{x} \le \frac{3\ell}{4}
\end{array}\right. ,
\end{align}
with the periodicity properties $\tilde F_y(\bar x+\frac{\ell}{2}, \bar p)=-\tilde F_y(\bar x, \bar p)$ and $\kappa_y(\bar x+\frac{\ell}{2}, \bar p)=\kappa_y(\bar x, \bar p)$, leading to
\begin{align}
&F_y(\hat x,\hat p) =\int_{-\ell/4}^{\ell/4} d\bar x \int_{-\pi/\ell}^{\pi/\ell} d\bar p  \, \tilde F_y(\bar x, \bar p) \nonumber \\
&\times i\left[ e^{-i\kappa_x(\bar x,\bar p)}  |\bar x+\frac{\ell}{2}, \bar p\rangle\langle\bar x, \bar p| -e^{i\kappa_x(\bar x,\bar p)} |\bar x, \bar p\rangle\langle\bar x+\frac{\ell}{2}, \bar p| \right]
 \nonumber \\
  &=\int_{-\ell/4}^{\ell/4} d\bar x \int_{-\pi/\ell}^{\pi/\ell} d\bar p  \, \tilde F_y(\bar x,\bar p) \hat\sigma_x(\bar x, \bar p) \equiv \hat\Gamma_y.
\end{align}
 where we defined
 \begin{align}
\hat\sigma_y(\bar x, \bar p) =&+ e^{+i\theta_+}|\bar x+\frac{\ell}{2}, \bar p\rangle\langle\bar x, \bar p|e^{-i\theta_-} \nonumber \\
&-ie^{+i\theta_-}|\bar x, \bar p\rangle\langle\bar x+\frac{\ell}{2}, \bar p|e^{-i\theta_+},
\end{align}
with $\theta_-(\bar x,\bar p)-\theta_+(\bar x,\bar p) =\kappa_y(\bar x,\bar p)=\kappa_x(\bar x,\bar p)$. Further on, consider the phase space operator (\ref{app:eq:DoubleFourier}), with $L=\ell/2$ and $L'=\ell$, yielding the matrix elements
\begin{align}
F_y(\bar x,& \bar p;\bar x', \bar p')=\frac{\ell}{2\pi}\sum_{\nu,\mu,n,m=-\infty}^{\infty}e^{i(\bar{p}'\mu-\bar{p}\nu)\ell} d_{n,m} \nonumber \\
&\times \bra{\bar x+\nu \ell}e^{4\pi i n \hat x/\ell-i \ell m \hat p/2}\ket{\bar x'+\mu \ell} \nonumber \\
=&\tilde F_y(\bar x,\bar p) \delta(\bar p-\bar p')  \nonumber \\
&\times\left[ie^{i\bar p\ell/2} \delta(\bar x-(\bar x'+\ell/2)) \Theta_1(\bar x)\right. \nonumber \\
&\left.+ie^{-i\bar p\ell/2} \delta(\bar x-(\bar x'-\ell/2)) \Theta_2(\bar x)\right],
\label{app:eq:MatrixEleY}
\end{align}
with 
\begin{align}
\tilde F_y(\bar x,\bar p)=&\sum_{n,m=-\infty}^{\infty} d_{2n+1,2m+1}  \nonumber \\
&\times e^{i\pi(n+m)}e^{2\pi i (2n+1) \bar x-i\bar p (2m+1) \ell/2}  .
\label{eq:MatrixEleY}
\end{align} 
We thus find that all operators of the form 
\begin{align} 
F_y(\hat x,\hat p)=&\sum_{n,m=-\infty}^{\infty} d_{2n+1,2m+1} \nonumber \\
&\times \hat D(\frac{\ell}{2}(2m+1),\frac{2\pi}{\ell}(2n+1)), 
\label{eq:FyObservable}
\end{align}
where we set $L=\ell/2$, $L'=\ell$ and $d_{n,m}=0$, for all even $n$ and $m$, can be expressed as $\hat\Gamma_y$ with $\zeta_y(\bar x,\bar p)=\tilde F_y(\bar x,\bar p)$ and $\kappa_y(\bar x,\bar p)=\bar p\ell/2$. An example of Eq. (\ref{eq:FyObservable}) is given by $\text{Re}(\hat Y)=\cos{(2\pi/\ell \hat x-\hat p\ell/2)}$, corresponding to the case where only $d_{1,1}=1/2$ and $d_{-1,-1}=-1/2$ are nonzero, which is equal to $\hat\Gamma_y$ with $\zeta_y(\bar x,\bar p)=F_y(\bar x,\bar p)=\cos{(2\pi/\ell \bar x-\bar p\ell/2)}$.


\begin{thebibliography}{99}
\bibliographystyle{unsrt}
\bibitem{Nielsen} M. A. Nielsen and I. L. Chuang, \textit{Quantum Computation and Quantum Information (Cambridge University Press, Cambridge, England, 2000)}.

\bibitem{Loock} S.L. Braunstein, and P. van Loock, Rev. Mod. Phys. \textbf{77}, 513 (2005).

\bibitem{Braunstein} S. L. Braunstein and H. J. Kimble, Phys. Rev. Lett. {\bf 80}, 869 (1998).

\bibitem{Furusawa1} A. Furusawa, J. L. S\o rensen, S. L.  Braunstein, C. A. Fuchs, H. J. Kimble and E. S. Polzik, Science {\bf 282}, 706 (1998).

\bibitem{Grosshans1} F. Grosshans, G. Van Assche, J. Wenger, R. Tualle-Brouri, N .J.  Cerf, and P. Grangier,  Nature {\bf 421}, 238-241
 (2003). 
 
 \bibitem{Grosshans2} F. Grosshans and P. Grangier, Phys. Rev. Lett. {\bf 88}, 057902 (2002). 
 
 \bibitem{Loyd} S. Lloyd and S. L. Braunstein, Phys. Rev. Lett. {\bf 82}, 1784 (1999). 
 
\bibitem{MBQCprl} N. C. Menicucci, P. van Loock, M. Gu, C. Weedbrook, T. C. Ralph, and M. A. Nielsen, Phys. Rev. Lett. \textbf{97}, 110501 (2006).
 
\bibitem{MBQC} M. Gu, C. Weedbrook, N. C. Menicucci, T. C. Ralph and P. van Loock, Phys. Rev. A \textbf{79}, 062318 (2009).

 \bibitem{Gottesman} D. Gottesman, A. Kitaev and J. Preskill, Phys. Rev. A {\bf 64}, 012310 (2001). 
 
 \bibitem{MenicucciFault} N. C. Menicucci, Phys. Rev. Lett. {\bf 112}, 120504 (2014). 
 
\bibitem{CHSH} J. F. Clauser, M. A. Horne, A. Shimony, and R. A. Holt, Phys. Rev. Lett. {\bf 23}, 880 (1969).

\bibitem{CVTomo} A. I. Lvovsky and M. G. Raymer, Rev. Mod. Phys. \textbf{81}, 299 (2009).

\bibitem{LG} A. J. Leggett and A. Garg, Phys. Rev. Lett. {\bf 54}, 857 (1985).

\bibitem{CabelloContext1} \'Adan Cabello, Phys. Rev. Lett. \textbf{101}, 210401 (2008).

\bibitem{Aharonov} Y. Aharonov, H. Pendleton, and A. Petersen, Int. J. Theo. Phys. \textbf{2}, 213 (1969). 

\bibitem{Clemens1} C. Gneiting and K. Hornberger, Phys. Rev. Lett. \textbf{106}, 210501 (2011).

\bibitem{ExpModVar} M. A. D. Carvalho, J. Ferraz, G. F. Borges, P. -L. de Assis, S. P\'adua, S. P. Walborn, Phys. Rev. A \textbf{86}, 032332 (2012).

\bibitem{mariana} M. R. Barros, O. J. Far\`ias, A. Keller, T. Coudreau, P. Milman, S. P. Walborn, Phys. Rev. A \textbf{92}, 022308 (2015).

\bibitem{Andreas} A. Ketterer, A. Keller, T. Coudreau and P. Milman, Phys. Rev. A \textbf{91}, 012106 (2015). 

\bibitem{AsadianCHSH} A. S. Arora and A. Asadian, Phys. Rev. A \textbf{92}, 062107 (2015).

\bibitem{Rabl} A. Asadian, C. Brukner and P. Rabl, Phys. Rev. Lett {\bf  112}, 190402 (2014).

\bibitem{Saulo} S. V. Moreira, A. Keller, T. Coudreau and P. Milman, Phys. Rev. A \textbf{92}, 062132 (2015).

\bibitem{CabelloContextCV} \'A. R. Plastino and A. Cabello, Phys. Rev. A \textbf{82}, 022114 (2010).

\bibitem{GuehneContext} A. Asadian, C. Budroni, F. E. S. Steinhoff, P. Rabl and O. G\"uhne, Phys. Rev. Lett. \textbf{114}, 250403 (2015).

\bibitem{AdrienContext} A. Laversanne-Finot, A. Ketterer, M. Barros, S. P. Walborn, A. Keller, T. Coudreau and P. Milman, arXiv:1512.03334v2 (2016).

\bibitem{Zak1967} J. Zak, Phys. Rev. Lett. \textbf{19}, 1385 (1967).

 \bibitem{errorboundsGKP} S. Glancy and E. Knill, Phys. Rev. A, \textbf{73}, 012325 (2006).
 
 \bibitem{Rabl1} O. Gittsovich, T. Moroder, A. Asadian, O. G{\"u}hne, and P. Rabl, Phys. Rev. A \textbf{91}, 022114 (2015).
 
 \bibitem{Pierre} P. Vernaz-Gris, A. Ketterer, A. Keller, S.P. Walborn, T. Coudreau and P. Milman, Phys. Rev. A {\bf 89}, 052311   (2014).
 
  \bibitem{Horodecki} P. Horodecki, Phys. Rev. A \textbf{67}, 060101 (2003).
 
 \bibitem{Tasca} D. S. Tasca, R. M. Gomes, F. Toscano, P. H. Souto Ribeiro, and S. P. Walborn, Phys. Rev. A \textbf{83} 052325 (2011).
 
 \bibitem{Talbot} O. J. Far\'ias, F. de Melo, P. Milman, S. P. Walborn, Phys. Rev. A \textbf{91}, 062328 (2015).
 
 \bibitem{Lantz} P.-A. Moreau, F. Devaux and E. Lantz, Phys. Rev. Lett \textbf{113}, 160401 (2014).

\bibitem{Camera1} R. S. Aspden, D. S. Tasca, R. W. Boyd and M. J. Padgett, New J. Phys. \textbf{15}, 073032 (2015).

\bibitem{FracFour1} H. M. Ozaktas, Z. Zalevsky, and M. A. Kutay, \textit{The Fractional Fourier Transform: with Applications in Optics and Signal Processing} (Wiley, New York, 2001).

\bibitem{FracFour2} D. S. Tasca, S. P. Walborn, P. H. Souto Ribeiro, and F. Toscano, Phys. Rev. A \textbf{78}, 010304(R) (2008).

\bibitem{interferometer1} S. Machado, P. Milman and S. P. Walborn, Phys. Rev. A {\bf 87}, 053834 (2013). 

\bibitem{interferometer2} M. Hor-Meyll, J. O. de Almeida, G. B. Lemos, P. H. Souto Ribeiro and S. P. Walborn, Phys. Rev. Lett. {\bf 112}, 053602 (2014). 

\end{thebibliography}
\end{document}